\documentclass[9pt,conference]{IEEEtran}
\usepackage{graphicx}
\usepackage{multirow}
\usepackage{multicol}
\usepackage{color}
\usepackage{url}
\usepackage{array}
\usepackage{algorithm}
\usepackage{algpseudocode}
\usepackage{cite}
\usepackage{amsmath}
\usepackage{mwe}
\usepackage{amssymb}
\usepackage[captionskip=0pt, caption=false]{subfig}
\definecolor{gray1}{gray}{0.7}
\definecolor{gray2}{gray}{0.98}
\definecolor{light-gray}{gray}{0.95}

\newcommand{\ignore}[1]{}
\newcommand{\redHL}[1]{\textcolor{red}{#1}}
\newcommand{\redfn}[1]{\textcolor{red}{\footnote{\textcolor{red}{#1}}}}

\usepackage{tikz}
\usetikzlibrary{shapes,arrows}
\usepackage{multirow}

\newcommand{\red}[1]{\textcolor{red}{ #1}}
\newcommand{\blue}[1]{\textcolor{blue}{ #1}}

\newcommand{\bluefn}[1]{\blue{\footnote{\blue{#1}}}}
\newcommand{\ziqing}[1]{{\color{blue}[Ziqing: #1]}}
\newcommand{\sachin}[1]{{\color{red}[Sachin: #1]}}
\newcommand{\husrev}[1]{{\color{green!60!black}[H\"usrev: #1]}}

\usepackage{pgfplots}
\usepgfplotslibrary{fillbetween}
\usepackage[eulergreek]{sansmath}
\pgfplotsset{
  tick label style = {font=\sansmath\sffamily},
  every axis label = {font=\sansmath\sffamily},
  legend style = {font=\sansmath\sffamily},
  label style = {font=\sansmath\sffamily}
}
\pgfplotsset{compat=newest}
\usepackage{complexity}
\usetikzlibrary{decorations.pathreplacing,calligraphy}
\usetikzlibrary{arrows.meta}

\usepackage{xcolor}
\usepackage{pgfplots}
\usepackage{tikz}
\usepackage{soul}
\usepackage{svg}
\definecolor{bblue}{HTML}{4F81BD}
\definecolor{rred}{HTML}{C0504D}
\definecolor{ggreen}{HTML}{9BBB59}
\definecolor{ppurple}{HTML}{9F4C7C}
\definecolor{oorange}{HTML}{FFAC1C}

\begin{document}
\title{3SAT on an All-to-All-Connected CMOS Ising Solver Chip}
\author{H{\"u}srev C{\i}lasun, Ziqing Zeng, Ramprasath S, Abhimanyu Kumar, Hao Lo, William Cho, \\
Chris H. Kim, Ulya R. Karpuzcu, and Sachin S. Sapatnekar}
\maketitle
\begin{abstract}
This work solves 3SAT, a classical NP-complete problem, on a CMOS-based Ising hardware chip with all-to-all connectivity. The paper addresses practical issues in going from algorithms to hardware.  It considers several degrees of freedom in mapping the 3SAT problem to the chip -- using multiple Ising formulations for 3SAT; exploring multiple strategies for decomposing large problems into subproblems that can be accommodated on the  Ising chip; and executing a sequence of these subproblems on CMOS hardware to obtain the solution to the larger problem.  These are evaluated within a software framework, and the results are used to identify the most promising formulations and decomposition techniques. These best approaches are then mapped to the all-to-all hardware, and the performance of 3SAT is evaluated on the chip. 
Experimental data shows that the deployed decomposition and mapping strategies impact SAT solution quality: without our methods, the CMOS hardware cannot achieve 3SAT solutions on SATLIB benchmarks.

\end{abstract}
\pagestyle{plain}
\section{Introduction}
\label{sec:Intro}

\noindent
Many combinatorial optimization problems (COPs), including NP-complete and NP-hard problems, can be solved using the quantum-inspired Ising model~\cite{lucas2014ising}, which originated from representations of magnetic interactions that settle to a minimum-energy state. Many COPs can be written in Ising form via quadratic unconstrained binary optimization~(QUBO) formulations, and then mapped to a network of coupled oscillators. As these oscillators settle to their minimum energy ground state, they solve the COP.  Ising hardware can potentially have better speed and energy than classical computers.

Many efforts have conceived or built Ising solvers in emerging technologies, e.g., quantum, spintronics, optics,
phase change devices, NEMS, and ferroelectrics. However, these substrates are not scalable or mass-manufacturable; some require prohibitively expensive cooling to a few Kelvin. In contrast, CMOS-based Ising solvers, which use  coupled ring oscillators (ROs), can make Ising computation practical, delivering high speed, low power consumption, accuracy, high integration density, portability, and mass-manufacturability. 

Many of today's Ising machines have limited connectivity: D-Wave's quantum-based solutions limit connectivity to 6--20 neighbors per oscillator~\cite{boothby2020next}; even many CMOS-based solutions~\cite{Yamaoka16,Ahmed21,moy20221} are limited to 4--8 nearest neighbors on a 2D oscillator mesh. The embedding problem of mapping the couplings in an Ising problem to this connectivity-limited structure requires spin replication: a six-variable problem with all-to-all interactions requires 18 spins on D-Wave's Chimera and 30 spins on the King's graph~\cite{Lo2023}. Replication weakens the strength of a spin, leading to suboptimal solutions~\cite{dwave-chainstrength}. 

Recent work breaks through these bottlenecks: in~\cite{Lo2023} a 65nm chip implements all-to-all (A2A) connectivity between 50 spins: its A2A connectivity makes it very powerful, equivalent to a locally connected architecture (e.g.,~\cite{Yamaoka16,Ahmed21,moy20221}) with thousands of spins~\cite{Tabi21}.

Even so, the problem of mapping general COPs to any Ising hardware substrate remains an open problem.  {\em First,} out of multiple QUBO formulations that are available for any COP, some may perform better on hardware than others.  {\em Second,} hardware engines must operate under a number of restrictions, e.g., (a)~the range of allowable coupling weights is limited; (b)~the solution accuracy can depend on the mapping strategy.  {\em Third,} since any hardware platform has limited capacity, large problems must be decomposed into smaller sub-problems, and the decomposition strategy impacts solution quality.

Thus, merely building a hardware substrate, is not enough; to move Ising computing closer to reality, it is essential to provide a complete solution from algorithms to hardware execution.  This work addresses these issues for 3SAT, a classical NP-complete problem.  We examine multiple choices of QUBO formulation, decomposition, and mapping strategies, and report results on actual CMOS hardware: a 65nm Ising chip with A2A connectivity~\cite{Lo2023}.  This exploration is essential to attain the optimal solution.  For example, depending on the formulation, the problem may require more or fewer spins, and more or fewer couplings; without a systematic evaluation, it is unclear which formulation delivers the best solution.  Similarly, decomposition and mapping strategies can significantly impact solution quality.

The contributions of this paper include: (1)~hardware-specific {\em evaluation of multiple mappings} from 3SAT to Ising models, (2)~rigorous methods for {\em variable pruning} through spin removal optimization, (3)~{\em scaling and local field oscillator optimizations} specifically for 3SAT, (4)~three novel {\em decomposers} to break large problems to subproblems that fit on the hardware, (5)~{\em hardware demonstration} of 3SAT benchmark instances on a CMOS Ising chip.

\ignore{
In recent years, many well-known optimization problems have been translated to the model of quadratic unconstrained binary optimization~(QUBO), the main motivation behind this is the QUBO models can be used as a problem specification for various quantum algorithms and quantum annealing. NP-complete computational problems require exponential solving time and energy even with the fastest known sequential algorithms on classical digital computers based on a von Neumann architecture. However, with QUBO formulation, physics-based Ising solvers can find competitive solutions to QUBO problems without the need for such complex algorithms translating into orders of magnitude better speed and energy\cite{moy20221}.

3-satisfiability~(3SAT) is a typical NP-complete computational problem~\cite{cormen2022introduction} 
is a Therefore, most of the existing 3SAT
~solving methodologies can be classified into classical digital computer-based and QUBO-based. In addition, methodologies other than the above two can only solve easy 3SAT~problems. For example, the DNA computer in ~\cite{braich2002solution} only solved 3SAT~problems with 20 variables and 24 clauses, where there are too many all-SAT solutions and easy to solve.

Translating the 3SAT~ problem into QUBO formulation leverages the nature of the quantum algorithms and quantum-inspired accelerators.

\husrev{QUBO essentially expresses its objective function as the QUBO \emph{Hamiltonian} that consists of a quadratic expression $\mathbf{x}^T Q \mathbf{x}$ with a real problem matrix $Q$ that contains problem-specific coupling coefficients and the binary solution vector $\mathbf{x}$. Finding a solution to the QUBO problem entails minimizing the QUBO Hamiltonian, which translates into finding the optimal binary vector $\mathbf{x}$. Although the problem formulation aims to find the global minima(minimum) that correspond(s) to ground Hamiltonian energies(energy), local minima(minimum) with higher Hamiltonians(Hamiltonian) may be obtained with a suboptimal solution.}

\begin{itemize}
\item 3SAT problem
\item Qubo/Ising formulation of 3SAT problem
\item Software Ising solver: Dwave
\item Hardware Ising solver: ring oscillator-based Ising machine
\item Our approach
\end{itemize}

3SAT problem can be transferred into a QUBO or Ising formulation. The existing QUBO/Ising software solver is not efficient to solve the corresponding problem. Therefore, we propose a ring oscillator-based Ising-solving platform to solve the 3SAT problem, which has high speed and energy efficiency.
}
\section{Solving Combinatorial Problems on Ising Machines}

\subsection{QUBO/Ising Problems and the Underlying Graph}
\label{sec:qubo_ising}

\noindent
A \textbf{QUBO problem} in $n$ variables is formulated as
\begin{equation} \label{eq:qubo}
    \hspace*{-3mm}
    \textstyle 
    \min_{\mathbf{x}} F(\mathbf{x}) = \mathbf{x}^T Q \mathbf{x} = \sum_{i=1}^n Q_{ii} x_i + \sum_{i=1}^n \sum_{j=1, j \not = i}^n Q_{ij} x_i x_j
\end{equation}
where $\mathbf{x} = [x_1, \cdots , x_n]^T \in \{ 0, 1 \}^n$ is a Boolean vector and $Q \in \mathbb{R}^{n \times n}$ is a real matrix; here, $Q_{ii}$ multiplies $x_i^2 = x_i$ for $x_i \in \{0, 1\}$. Using $x_i = (s_i + 1)/2$ to transform each Boolean variable $x_i$ to a spin variable $s_i \in \{-1, +1\}$, the isomorphic \textbf{Ising formulation} is 
\begin{equation} \label{eq:ising}
    \textstyle
    \min_{\mathbf{s}} F(\mathbf{s}) = \sum_i h_i s_i + \sum_{i=1}^n \sum_{j=1, j \neq i}^n J_{ij}s_i s_j
\end{equation}
where $\mathbf{s} = [s_1, \cdots s_n]^T \in \{+1,-1\}^n$, $J \in \mathbb{R}^{n \times n}$ is an upper triangular real matrix, and $\mathbf{h} = [h_1, \cdots, h_n]^T \in \mathbb{R}^n$ is a real vector, where
$h_i = Q_{ii}/2 + \sum_{j=1}^n (Q_{ij} + Q_{ji})/4$ and $J_{ij}=Q_{ij}/4$. 
In both formulations, the objective functions characterize the \textit{Hamiltonian}, which represents the energy measure of the system.

The \textbf{graph representation} of the Ising formulation associates each variable $s_i$ with a vertex $i$ with weight $h_i$, with coupled vertices $i$ and $j$ connected by an undirected edge of weight $(J_{ij}+J_{ji})$. We will use this graph interchangeably with the optimization formulation.

\subsection{A CMOS-based Ising Hardware Accelerator}
\label{sec:cmos_hardware}

\begin{figure}[bt]
    \centering
    \subfloat[]{
        \includegraphics[width=.55\linewidth]{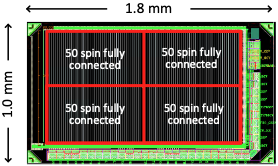}
        \includegraphics[width=.28\linewidth]{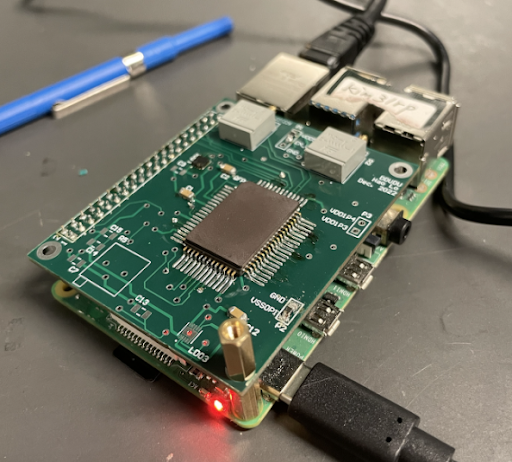}
    }\\
    \subfloat[]{
        \vspace{0em}
        \includegraphics[width=0.7\linewidth]{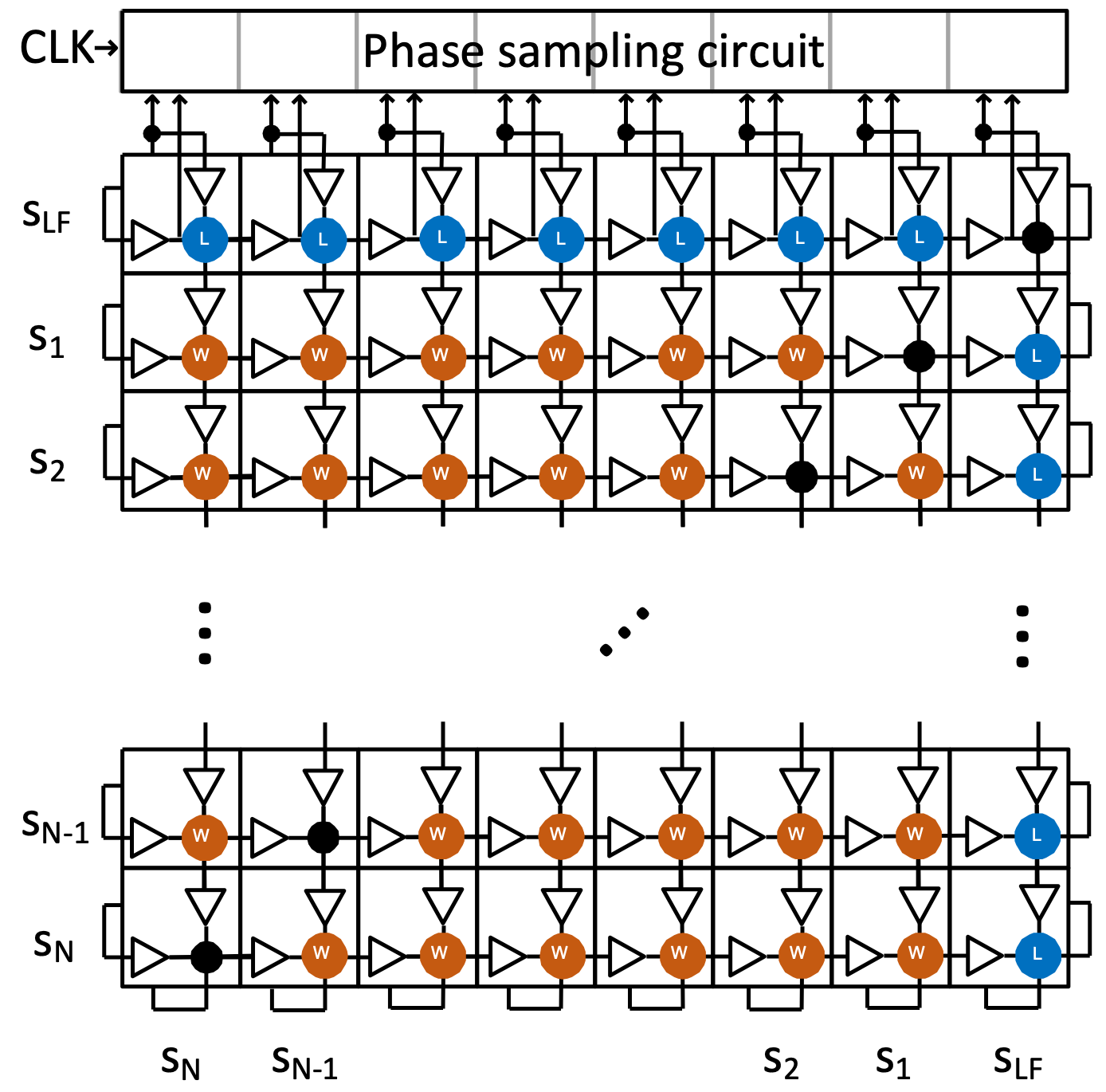}
        \vspace{0em}
    }
\caption{\cite{Lo2023} (a) Chip layout and packaged chip soldered on a carrier board, attached to a Raspberry Pi. (b) All-to-all connected array of CMOS ROs.}
\label{fig:architecture}
\vspace{-6mm}
\end{figure}

\noindent
Fig.~\ref{fig:architecture} illustrates our hardware engine with an A2A architecture~\cite{Lo2023}, comprising (N$+1$) horizontal oscillators and (N$+1$) vertical oscillators. Each horizontal oscillator is shorted with the corresponding vertical oscillator, as shown by the black dots on the diagonal, so that the horizontal and vertical oscillators form a single physical oscillator carrying the same phase information.  The spin variable associated with an oscillator corresponds to its phase.  The paired oscillators denoted as s$_{\rm LF}$ are phase-locked and serve as the timing reference for the entire array, with spin value of $s_{\rm LF}$ fixed at $+1$; for other spins in the array whereas the spin values of $s_i$ are either $+1$ or $-1$, depending on whether the phase is the same as/opposite to $s_{\rm LF}$. 
 
Since $s_{\rm LF}=+1$ serves as the fixed reference, the coupling term between oscillator s$_{\rm LF}$ and s$_i$ is $J_{i,{\rm LF}} s_i s_{\rm LF} = J_{i,{\rm LF}} s_i$. This becomes the local field term $h_i$ (= $J_{i,{\rm LF}}$) in the Ising Hamiltonian equation. In Fig.~\ref{fig:architecture}, the coupling circuits along the bottom and right edges of the array, denoted as L, implement the $h_i$ weights. The intersection between spins $s_i$  and $s_j$, denoted as W, implement the coupling weights $J_{ij}$ between spins $s_i$ and $s_j$.  The coupling circuits are implemented with transmission gates~\cite{Lo2023}.

The weight $J_{ij}$ can be implemented by programming two locations -- row $i$, column $j$, and row $j$, column $i$ -- in the array and the weight is the sum of weights in these locations. Since each coupling site can implement a weight of up to $\pm 7$, $J_{ij} \in \{-14,+14\}$.

The phase sampling block in Fig.~\ref{fig:architecture} samples each RO 8 times in each cycle of the RO to generate an 8-bit binary result that is read out to determine the phase of the RO by majority voting~\cite{Lo2023}.

\section{Degrees of Freedom in A2A Hardware Mapping}
\label{ssec:mapping_Ising}

\noindent
To solve a COP in Ising form on the A2A hardware of Section~\ref{sec:cmos_hardware},
the process of mapping the Ising matrix and local field to the hardware must work within the hardware limitations. Since the coupling values $J_{ij}$ must be integers in the range $[-14,+14]$, smaller coupling weights of the problem may have to be \textbf{upscaled} while larger weights must be \textbf{downscaled} to lie in $[-14,14]$.  This scaling step may result in non-integer values; these are rounded to an integer.  

Conflicting considerations must be balanced during scaling: 
(1)~The device accuracy is proportional to the coupling strength and therefore large scaling values are preferable. 
(2)~If most of the weights are low in magnitude and only a few are high, it is possible that the lower weights will be zeroed out during downscaling.  To avoid this, some coupling weights may be scaled beyond the dynamic range of the device and then clamped to the nearest extreme limit of the weight range. Excessive scaling/clamping may alter the coupling matrix so greatly that its solution departs from that of the original problem.
Thus, scaling introduces trade-offs: we address them by dynamically reconfiguring the weight resolution of each spin. 

A similar trade-off exists for the local field oscillators. The device allows an arbitrary number of spins to be configured as local field ROs (which implement $h_i$), while the remaining spins are configured to maintain pairwise coupling ($J_{ij}$ values). Increasing the number of Local Field Ring Oscillators (LFROs) can increase the dynamic range of the $h$ coefficients: for example, a single LFRO can allow a coupling weight in the range $[-14,+14]$; if we perform {\em spin merging}, where two LFROs are used (and coupled tightly) to represent a single spin, a weight range of $[-56,56]$ (as explained in the next paragraph) is allowable. In particular, if the effective dynamic range of the $h_i$s is larger than that of the $J_{ij}$s, then increasing the LFRO count can allow higher ranges for coupling values with less truncation, which is seen to result in improved accuracy. However, this also translates into fewer spin variables being available for problem mapping.

\begin{figure}[t]
\vspace{0em}
\centering
\includegraphics[width=\linewidth]{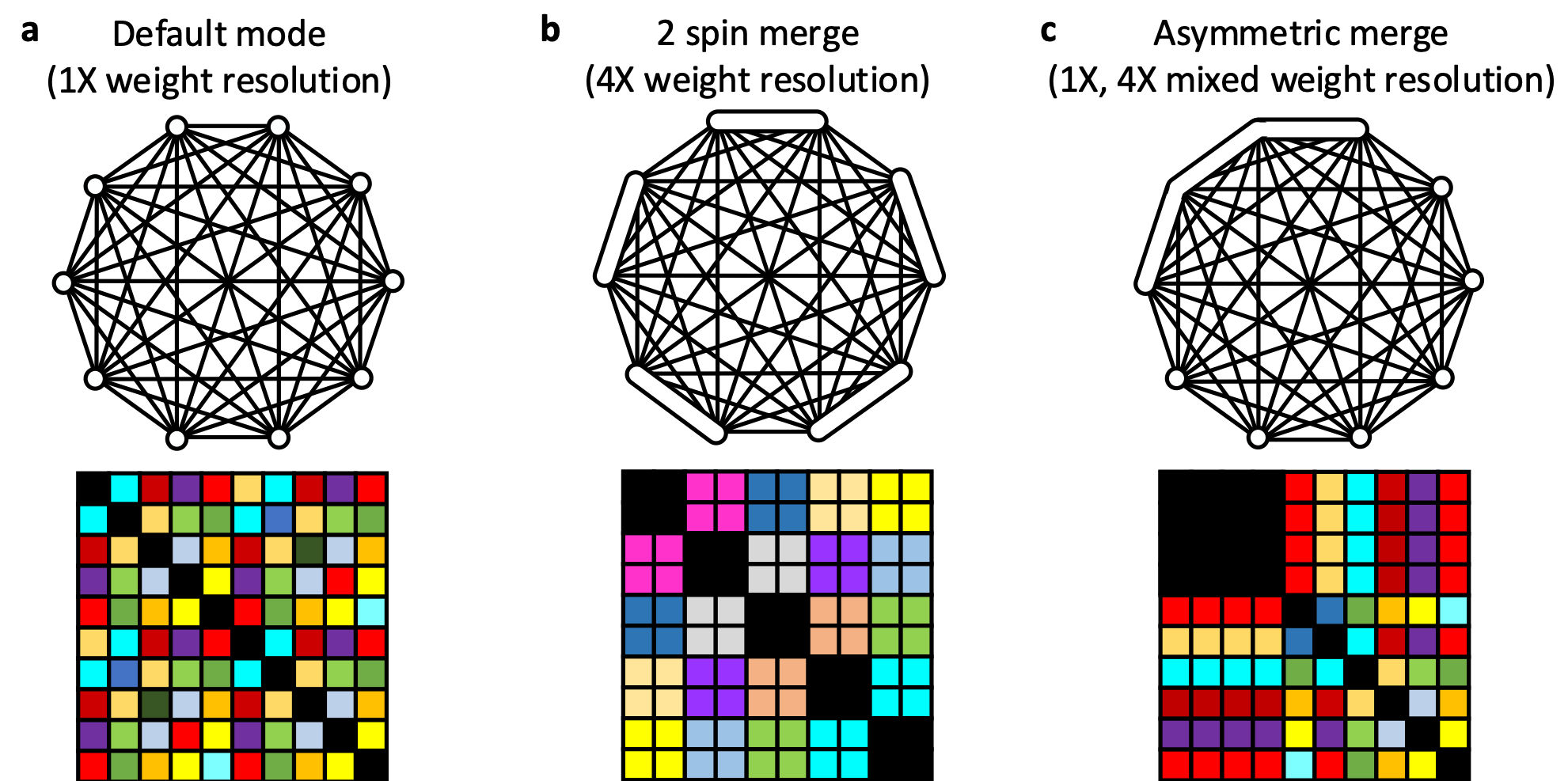}
\vspace{-6mm}
\caption{Illustrating spin-merging in the all-to-all array~\cite{Lo2023}.}
\label{fig:spinmerging}
\vspace{-6mm}
\end{figure}

We illustrate spin merging in Fig.~\ref{fig:spinmerging}~\cite{Lo2023} where a 10-spin hardware example is configured to a 10-spin default mode, a 5-spin 4× resolution mode, and a 6-spin asymmetric resolution mode, respectively. The graph (upper row) and the corresponding hardware mapping (lower row) are shown for each configuration. 
In the latter, the black weight cells connect the vertical and horizontal oscillators, and the coupling cells are color-coded according to coupling strength.
When two spins are merged (middle figure), coupling sites (each with weights up to $\pm 7$) lie on two $2 \times 2$ off-diagonal arrays. This allows coupling of $[-28,28]$ at each site and $J_{ij} \in [-56,+56]$. Thus, weight resolution can be traded off with the number of available spins.

\section{Formulating 3SAT for an Ising solver} 
\label{sec:Formulations}


\noindent
The Boolean satisfiability problem seeks to find an assignment of input variables for which a Boolean function evaluates to logic 1.  
The 3SAT problem was the ``original'' NP-complete problem,
and 3SAT reduces to any other NP-complete problem through a polynomial-time transformation~\cite{Karp72}.  Therefore 3SAT~is representative in the sense that its solutions can be used to obtain solutions to all NP-complete problems. Such problems span from practical challanges such as Traveling Salesman Problem \cite{Kabadi2007} to numerous applications in electronic design automation~\cite{Wang09}.

A 3SAT~instance in Conjuctive Normal Form (CNF) is a conjunction of clauses, i.e., $f(x_1, \cdots, x_n) = C_1 \land C_2 \land \cdots \land C_m$, where $X = \{x_1, \cdots, x_n\}$ is a set of $N$ Boolean variables. Each clause $C_i = l_1 \lor l_2 \lor l_3$ is a disjunction of at most three literals $l_1, l_2, l_3 \subset X \cup \neg X$. A SAT~formula $f$ is  
\textit{satisfiable} if there exists a set of Boolean assignments from $\{0,1\}$ on each variable in $X$ that can be substituted such that $f(x_1, \cdots, x_n) = 1$; any combination of such variables, if it exists, is called a \textit{satisfying assignment}. 

Max-3SAT is a variant of the 3SAT~problem that maximizes the number of clauses satisfied. If the solution of Max-3SAT says that all $N$ clauses are satisfiable, then the corresponding 3SAT problem is satisfiable~\cite{max3sat}. The Max-3SAT problem can be formulated in QUBO/Ising Hamiltonians using multiple formulations, which we will describe next. We show these formulations in QUBO form; the spin formulation can be found as shown in Section~\ref{sec:qubo_ising}. The superscript against the name of each formulation provides the number of spins in the formulation as a function of $n$ and $m$.

\subsection{The MIS\textsuperscript{$3m$} Formulation} 
\label{ssec:MIS}

\noindent
The maximal independent set (MIS) formulation~\cite{Dasgupta08} establishes a graph by assigning a QUBO variable (vertex) for each literal. For each clause, the three literals (vertices) are connected to each other, forming a ``triangle'' of edges. The literals of different clauses interact via conflict edges, which are formed if any two literals are negated versions of the same original Boolean variable. The triangle encourages the QUBO formulation to move to a new clause once a specific clause is satisfied, thus encouraging Max-3SAT, and penalizes conflicting assignments for the same variable. For a problem with $m$ clauses, this formulation requires $3m$ variables.  

\ignore{
Fig.~\ref{fig:3SAT2MIS} shows an example of mapping a Max-3SAT problem to its MIS\textsuperscript{$3m$} equivalent.

\begin{figure}[ht!]
\centering
\includegraphics[width=0.8\linewidth]{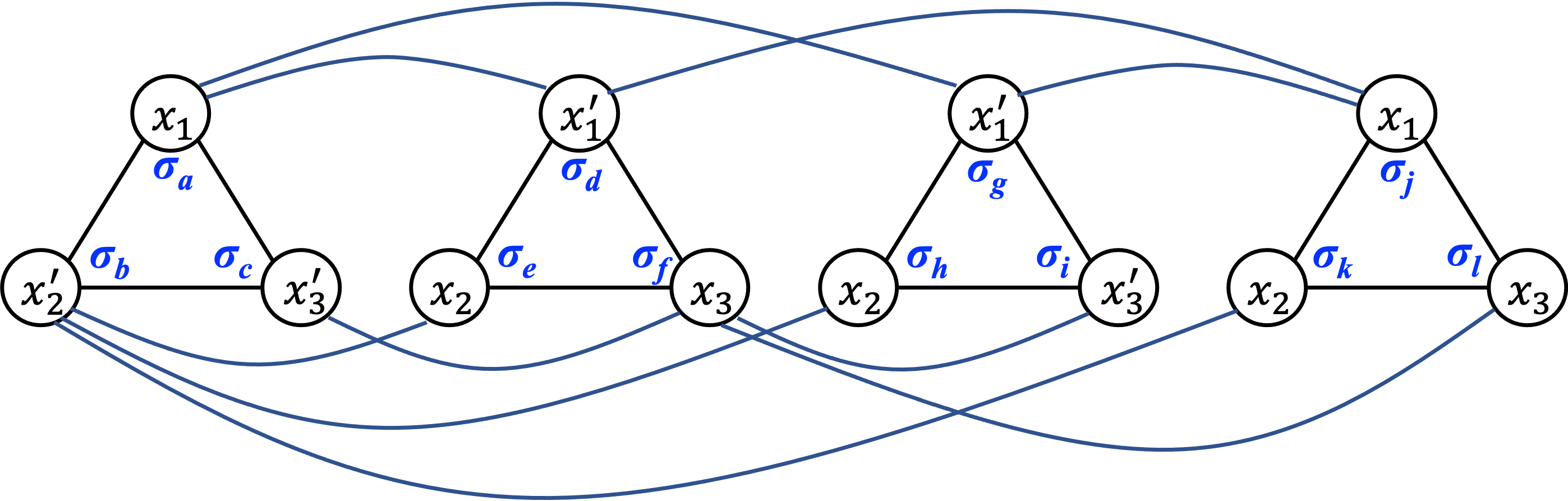}
\caption{The 3SAT instance $f = (x_1 \vee x_2' \vee x_3')(x_1' \vee x_2 \vee x_3)(x_1' \vee x_2 \vee x_3')(x_1 \vee x_2 \vee x_4)$
as an MIS problem.
}
\label{fig:3SAT2MIS}
\end{figure}

\begin{figure}[ht!]
    \centering
    \begin{tikzpicture}
        \node[shape=circle,draw=black] (x1) at (0,0) {$x_1$};
        \node[shape=circle,draw=black] (x2) at (1,0) {$x_2$};
        \node[shape=circle,draw=black] (x3) at (2,0) {$x_3$};
        \node[shape=circle,draw=black] (x4) at (3,0) {$x_4$};
        \node[shape=circle,draw=black] (x5) at (0,1.5) {$x_5$};
        \node[shape=circle,draw=black] (x6) at (1,1.5) {$x_6$};
        \node[shape=circle,draw=black] (x7) at (2,1.5) {$x_7$};
        \node[shape=circle,draw=black] (x8) at (3,1.5) {$x_{8}$};
\draw [decorate,
    decoration = {calligraphic brace}] (-0.5,1.2) --  (-0.5,1.8) node[pos=0.5,left=1pt,black]{Ancillary (Clause) variables};
\draw [decorate,
    decoration = {calligraphic brace}] (-0.3,0.35) --  (-0.3,1.1) node[pos=0.5,left=1pt,black]{Ancillary-to-Original variable edges};
\draw [decorate,
    decoration = {calligraphic brace}] (-0.5,-0.30) --  (-0.5,0.25) node[pos=0.5,left=1pt,black]{Original variables};
\draw [decorate,
    decoration = {calligraphic brace}] (-0.2,-1.00) --  (-0.2,-0.40) node[pos=0.5,left=1pt,black]{Original-to-Original variable edges};
        \path [-] (x1) edge node[left] {} (x5);
        \path [-] (x2) edge node[left] {} (x5);
        \path [-] (x3) edge node[left] {} (x5);
        \path [-] (x1) edge node[left] {} (x6);
        \path [-] (x2) edge node[left] {} (x6);
        \path [-] (x3) edge node[left] {} (x6);
        \path [-] (x1) edge node[left] {} (x7);
        \path [-] (x2) edge node[left] {} (x7);
        \path [-] (x3) edge node[left] {} (x7);
        \path [-] (x1) edge node[left] {} (x8);
        \path [-] (x2) edge node[left] {} (x8);
        \path [-] (x4) edge node[left] {} (x8);
        \path [-] (x1) edge[bend right=50] node[left] {} (x2);
        \path [-] (x1) edge[bend right=60] node[left] {} (x3);
        \path [-] (x2) edge[bend right=50] node[left] {} (x3);
        \path [-] (x1) edge[bend right=70] node[left] {} (x4);
        \path [-] (x2) edge[bend right=70] node[left] {} (x4);
    \end{tikzpicture}
    \caption{The 3SAT instance $f = (x_1 \vee x_2' \vee x_3')(x_1' \vee x_2 \vee x_3)(x_1' \vee x_2 \vee x_3')(x_1 \vee x_2 \vee x_4) 
    $ in Chancellor\textsuperscript{$n{+}m$}
    formulation.}
    \label{fig:3SAT2IDT}
\end{figure}
}

This formulation was translated into QUBO form in~\cite{choi2010adiabatic}, using up to $3m$ QUBO variables, one for each vertex in the MIS formulation.  Given an instance $C_1 \lor C_2 \lor \dots C_m = (l_1 \lor l_2 \lor l_3) \land (l_4 \lor l_5 \lor l_6) \land \dots \land (l_{3m-2} \lor l_{3m-1} \lor l_{3m})$, the QUBO Hamiltonian is:
\begin{multline} \nonumber 
    \textstyle -\sum_{i=1}^{3m} l_i  + 
            2\left(\sum_{i=0}^{m-1} \sum_{j<k \in [1,3]} l_{3i+j} l_{3i+k} + \sum_{i,j|l_i=\neg l_j} l_i l_j \right) 
\end{multline}
where $\mathbf{l}$ is the vector of all literals in the expression.  The minimized Hamiltonian represents the solution of the Max-3SAT~problem.

The literals, $l_i$, have a many-to-one mapping to the original Boolean variables. If the literal values provide conflicting assignments to the Boolean variables, a majority vote is used to assign the value.

\ignore{
\subsection{$k$-SAT~and 3SAT}
\label{ssec:SAT}
$k$-satisfiability ($k$-SAT) problem is one of the most fundamental problems in computational complexity. Despite being asymptotically hard, the attached significance is partially due to $k$-SAT~being NP-complete \cite{cook1971complexity}, which guarantees a polynomial-time reduction between NP-complete problems. Another polynomial reduction is available to reduce the degree of $k$ to 3 when $k>3$, resulting in 3SAT~problems. Therefore, 3SAT~is representative in the sense that its solutions can be used to obtain solutions to all NP-complete problems. Such problems span from practical challanges such as Traveling Salesman Problem \cite{Kabadi2007} to industrial applications such as Electronic Design Automation (EDA) \cite{998437}.

A $k$-SAT~problem (\textit{i.e.} instance) in Conjuctive Normal Form (CNF) can be represented as a conjunction of clauses as $f(x_1,x_2,\dots,x_n)=C_1\land C_2 \land \dots \land C_m$, $C_m$ being the $m$-th clause for $N$ Boolean variables $X=x_1,x_2,\dots,x_n$. A clause $C$ consists of $k$ literals $l_1,l_2,\dots,l_k$ that are arbitrarily sampled from $X\cup \neg X$, and can be defined as $C=l_1 \lor l_2 \lor \dots \lor l_k$. A SAT~formula $f$ is only can be \textit{satisfiable} if there exists a set Boolean assignments from $\{\textrm{True},\textrm{False}\}$ on $X$ that can be substituted to $f$ to render it $True$. Consecutively, the combination(s) of such variables is called \textit{satisfying assignment(s)}, if exists. A Max-3SAT~problem is a variant of the 3SAT~problem which tries to maximize the number of clauses satisfied, whose optimal solution is equivalent to 3SAT\cite{max3sat}. Max-3SAT~can be formulated in QUBO and Ising Hamiltonians using multiple formulations as follows:

\paragraph{Maximal Independent Set (MIS)} Following the formulations in \cite{choi2010adiabatic,lucas2014ising}, MIS formulation requires $3m$ QUBO variables (for $m$ clauses in the 3SAT~instance $C_1 \lor C_2 \lor \dots C_m = (l_1 \lor l_2 \lor l_3) \land (l_4 \lor l_5 \lor l_6) \land \dots \land (l_{3m-2} \lor l_{3m-1} \lor l_{3m})$ to establish the following objective function:

\begin{multline} \nonumber 
    \min_x \sum_{i=1}^{3m} x_ix_i + 2\left(\sum_{i=1}^{m} \sum_{j<k \in [0,2]}\hspace{-.3cm}x_{3i+j}x_{3i+k}+\hspace{-.3cm}\sum_{i,j|l_i=\neg l_j} \hspace{-.3cm}x_i x_j \right)
\end{multline}

The argument of \eqref{eq:mis} represents the solution of Max-3SAT~problem. MIS formulation establishes a graph by assigning a QUBO variable for each literal. Each literal is connected to all other vertices in the same clause in a triangle fashion. The literals of different clauses interact via conflict edges, which are formed if any two literals are negated versions of the same original Boolean variable. This way the QUBO formulation penalizes the solutions which do not satisfy all clauses, as well as the ones with conflicting assignments for the same variable. 

The solution of the MIS formulation cannot be directly inferred as the QUBO variables represent distinct literals, not original Boolean variables. The solution interpretation is based on a walk through the list of literals. If any QUBO variable is \texttt{True}, then the solution for the respective Boolean variable is assigned based on whether the QUBO variable is negated. In a satisfying assignment, negated literals of the same Boolean variable cannot be in the solution set. If the solution set lacks a certain Boolean variable, then such a variable can either be assigned as \texttt{True} or \texttt{False}.
}

\subsection{An ILP\textsuperscript{$n{+}2m$} Formulation} 
\label{ssec:ILP}

\noindent
Using the notation introduced above, we propose a new formulation representing the $i^{\rm th}$  clause by the Boolean inequality:
\begin{equation*}
l_{3i+1} + l_{3i+2} + l_{3i+3} \geq 1
\end{equation*}
If the literal $l$ corresponds to variable $x$ in true form, then $l = x$; else $l = 1- x$ if the literal is negated. The Max-3SAT problem can be represented as an integer linear program (ILP) that finds a feasible solution under these inequality constraints.  Using a slack variable $s$, each inequality constraint is transformed to an equality constraint $l_{3i+1} + l_{3i+2} + l_{3i+3} - s - 1 = 0$, where $s \in \{0, 1, 2\}$, depending on whether one, two, or all three literals are 1.  Encoding $s = 2s_{i,1} + s_{i,0}$, where $s_{i,1}$ and $s_{i,0}$ are binary variables, the equality constraint now contains all binary variables. To ensure this equality for each clause, the Hamiltonian minimizes the following sum of quadratics: 
\begin{equation*}
\textstyle \sum_{i=0}^{m-1} \left ( l_{3i+1} + l_{3i+2} + l_{3i+3} - 2s_{i,1} - s_{i,0} - 1  \right )^2
\label{eq:ilp}
\end{equation*}
For example, clause $(x_a \lor x_b' \lor x_c)$ is encoded as $x_a + (1 - x_b) + x_c - 2s_1 - s_0 - 1 = 0$, and contribution to the Hamiltonian is the square of the left hand side.
An instance with $n$ variables and $m$ clauses has $n+2m$ QUBO variables, including two ancillary slack variables for each of $m$ clauses. Typically, $n < m$, this has fewer Ising variables than MIS, but the range of weights is higher and the connectivity is denser.
Unlike the MIS$^{3m}$ formulation, where multiple variables represent the same literal and could potentially have contradictory assignments, there is a 1--1 correspondence between the first $n$  QUBO/Ising variables and the Boolean Max-3SAT variables. 


\subsection{The Chancellor\textsuperscript{$n{+}m$} Formulation} 
\label{ssec:IDT}

\noindent
The formulation in \cite{chancellor2016direct} maps an $n$-variable $m$-clause instance using $n+m$ QUBO/Ising variables. The formulation has a 1--1 correspondence between the $n$ SAT variables and the first $n$ QUBO/Ising variables, and adds one ancillary variable for each of the $m$ clauses.  Denoting the SAT variables as $x_1, \cdots, x_n$ and the ancillary variables as $x_{n+1}, \cdots, x_{n+m}$, the overall Hamiltonian is:
\begin{multline} 
     \hspace{-6mm}
     \textstyle \sum_{i=0}^{m-1} \left( -(l_{n+i}\hspace{-.1cm}+\hspace{-.1cm}1)(l_{3i}\hspace{-.1cm}+\hspace{-.1cm}l_{3i+1}\hspace{-.1cm}+\hspace{-.1cm}l_{3i+2} )+2l_{n+i}+ 
     \textstyle \sum_{j<k \in [0,2]}l_{3i+j}l_{3i+k}\right)    \nonumber 
\end{multline} 
As in ILP$^{n+2m}$, for a literal $l$ in true form, $l = x$; else $l = 1- x$. 

\subsection{The N\"u{\ss}lein\textsuperscript{$2n{+}m$} Formulation}

\noindent
The N\"u{\ss}lein\textsuperscript{$2n{+}m$} formulation~\cite{nusslein2023solving} maximizes the number of satisfied clauses by making the Hamiltonian equal to the negative of the number of the satisfied clauses. For this purpose, a dual of each of the $n$ original variables is designated to obtain the variable pairs $x_i, x_{i+1}$ that correspond to the $i^{\rm th}$ 3SAT variable.  These are one-hot-encoded to 10 if the 3SAT variable is true, and 01 if it is false. Additionally, one ancillary variable is designated for each of $m$ clauses, leading to $2n+m$ variables.  
The QUBO weights are assigned according to the following Hamiltonian:
\begin{align}
&\textstyle {-}\sum_{i=1}^{2n} R(x_i) x_i^2 + 2 \sum_{i=2n+1}^{2n+m} x_i^2 + 
    \sum_{i=1 | (i \hspace{-.2cm} \mod 2) \not = 0}^{2n-1} (m+1) x_i x_{i+1} \nonumber \\
& + \textstyle \sum_{i=1}^{2n} \sum_{j=1}^{2n} R(x_i,x_j) x_i x_j 
    - \sum_{i=1}^{2n} \sum_{j=2n|x_i \in c_{j-2n}}^{2n+m} x_i x_j 
\nonumber
\end{align} 
where $R(x_i)$ is the number of clauses that contain $x_i$,
and $R(x_i,x_j)$ is similarly defined as the number of clauses such that contain both $x_i$ and $x_j$.
This Hamiltonian aims to make the energy contribution of each satisfied clause $-1$ (and each unsatisfied clause $0$). The formulation then ensures that each variable is rewarded if it satisfies a clause (first term), and the local field coefficient of the ancillary variable of each clause is assigned to 2 (second term). Assignments of both a Boolean variable and its complement to 1 are penalized to ensure consistency (third term); the case where both are 0 effectively means that the variable is a don't care.

\subsection{The N\"u{\ss}lein\textsuperscript{$n{+}m$} Formulation}

\noindent
Although Chancellor\textsuperscript{$n{+}m$} produces a relatively low number of QUBO variables than MIS\textsuperscript{$3m$} and N\"u{\ss}lein\textsuperscript{$2n{+}m$}, it uses relatively more coupling weights.
The N\"u{\ss}lein\textsuperscript{$n{+}m$} formulation~\cite{nusslein2023solving} has a lower number of nonzero couplings between QUBO/Ising variables.

N\"u{\ss}lein\textsuperscript{$n{+}m$} is based on four different clause literal negation patterns (provided in \cite{nusslein2023solving}) for Max-3SAT (where each pattern corresponds to the number of negated literals in the clauses). The formulation then consists of constructing (or \textit{updating}) the Hamiltonian, clause by clause. Each pattern ensures that post(pre)-update Hamiltonian $H^{+}$($H^{*}$) satisfies $H^{+}=H^{*}$ if the immediate clause is satisfied, and $H^{+}=H^{*}+1$ otherwise. Based on this rule and the negation pattern of each clause, the Hamiltonian is updated iteratively for each clause. An advantage is the possibility of reusing clause variables; in the worst case, the number of clause variables is $m$, as in Chancellor\textsuperscript{$n{+}m$}.



\ignore{
\section{Solution Techniques for Ising Formulations}
\label{sec:solution_Ising}

\subsection{Existing Hybrid Solvers}
\label{ssec:hybrid_solver}

\noindent
QUBO/Ising solvers are used to find the optimal solution given a QUBO~\ref{eq:qubo} or Ising~\ref{eq:ising} Hamiltonian. Apart from some software-based solvers, other existing computing methodologies involving Quantum or Neuromorphic hardware are declared to be hybrid. The reason for using the hybrid method is the physical limit of the hardware, which is not able to cover all spins or variables in most standard benchmarks. A typical example is \textbf{D-Wave-Hybrid}~\cite{dwave-hybrid}, a hybrid classical/quantum solver~\cite{raymond2023hybrid} including a two-level approach, which is widely used in many QUBO/Ising optimizations~\cite{pastorello2019quantum,asproni2020accuracy}. Therefore, we will introduce some existing QUBO solvers from the software level and the hardware level.

On the software level, the classical software-based implementation of D-Wave-hybrid is called \textbf{QbSolv}~\cite{qbsolv}, which is a decomposer-involved solver that finds a minimum value of a large QUBO problem by splitting it into Sub-QUBO problems. It includes global Tabu searches, decomposer, and local Tabu searches. In the global Tabu search, the Qbsolve solver only searches for a global initiation point to start the local optimizations. Then, the decomposer generates the sub-QUBO problem based on the energy impact~(flip energy) of the spins and selects spins with the highest energy impact~(flip energy deviation). Sub-QUBO problems are generated by freezing all other unselected spins. The size  of the sub-QUBO problems can be customized. The sub-QUBO can be solved by user-defined software algorithms such as local Tabu search. After all sub-QUBO problems are solved, the overall solution will be passed to the global Tabu search for the next iteration. 

On the hardware level, \textbf{quantum processing unit~(QPU)} such as D-Wave 2000Q can be used to solve the sub-QUBO problems in \textbf{QbSolv}, executing local optimization on a non-von Neumann hardware to make it a hybrid quantum solver, mentioned as D-Wave-Hybrid. Specifically, D-Wave QPU currently does not support an all-to-all connection graph, they use one of three following simplified qubit-coupling topologies: Chimera~\cite{boothby2016fast}, Pegasus~\cite{dattani2019pegasus}, and Zephyr~\cite{boothby2020next,boothby2021zephyr}. Therefore, the D-Wave quantum-annealing system is very good at exploring diverse regions of the state space but its limited precision constrains its ability to get to an exact minimum quickly\cite{king2015benchmarking}.
\ziqing{should I add the description for the above 3 formulations and all-to-all connection?}

\
\ignore{Another typical non-von Neumann QUBO solver is \textbf{IBM’s Neurosynaptic
TrueNorth System}~\cite{merolla2014million},\redfn{This reference is very puzzling. TrueNorth is not a QUBO solver, as far as I know: in fact, I could not find ``QUBO'' when I searched their paper.  There are a few papers that have tried to implement QUBO on true north (\url{https://www.google.com/search?client=firefox-b-1-d&q=true+north+qubo}, but none of them seems to have many citations.}\bluefn{s} the architecture of which is inspired by biological neural systems. NS16e is a 16-chip system where 16 are incorporated together for reconfigurable Neuromorphic programming. Synaptic weights are signed integers from -127 to 128 and each neuron’s and synapse’s states are updated every millisecond. Four different kinds of QUBO problems with $8\times8$ and $16\times16$ input weight matrix are tested on the NS16e TrueNorth System to show promising accuracy and power consumption~\cite{alom2017quadratic}. Although the NS16e uses an all-to-all connection graph as input formulation, the $16\times16$ dimension would be a barrier in solving large benchmarks.}
\ziqing{Paragraph about TrueNorth is removed, plan to add a paragraph about photonic Ising machine about~\cite{inagaki2016coherent,pierangeli2019large}}

\husrev{Should we mention Fujitsu's digital annealer as well: https://www.fujitsu.com/global/services/business-services/digital-annealer/}\sachin{Incorporated into the beginning of Section~\ref{sec:cmos_hardware}; may be moved to Intro later.}

\ignore{
\subsection{Ring Oscillator-Based All-to-All Ising Solver}
\sachin{Most likely, this will go away, since it has been incorporated into an earlier section. We need to find a home for the previous subsection.}

Different from other D-Wave processors that use simplified qubit-coupling topologies, ring oscillator~(RO)-based Ising solver~\cite{moy20221} implements all-to-all couplings with the coupling of ring oscillators on a chip fabricated in a standard 1.2V, 65-nm CMOS technology. Each RO, consisting of 7 stages of inverters, acts as a Qubit which is coupled to all other ROs with transmission gates. Each coupling value $J_{ij}$ or $h_{i}$ is implemented as the strength of 14 transmission gates which are programmable for the coupling ranges from -14 to +14 in a step of 1. Specifically, One RO can be assumed as the reference then all couplings to the reference correspond to $h_{i}$. All couplings do not involve reference corresponding to $J_{ij}$. Solving an Ising problem on the chip includes the following steps: Setting up the transmission gates based on coupling values, waiting some time for the ROs array to synchronize, and sampling the phase of all ROs. The binary solution of each Ising spin is determined by the sampling phase of the corresponding RO. In one cycle of the reference RO, there is a sampling block that samples each RO 8 times to generate an 8-bit binary result as the phase information.  
The Ring oscillator-based all-to-all Ising solver is very fast and energy-efficient to solve the QUBO/Ising problem. However, due to the unstable and noise of analog signals, there is no guarantee that the RO array can always solve the Ising machine. And the solving effectiveness is problem-dependent. For example, the random problem-solving rate decrease with the decrease in coupling density. Therefore, to achieve higher solving accuracy, we need to run the RO array many times and choose the solution with optimal Hamiltonian energy.

The hardware solver we use now is the

Background of Ising optimization and hardware \redHL{Ziqing}
\begin{itemize}
\item Hardware Ising solver: Ring oscillator-based Ising machine
\item Software Ising solver: D-Wave QBsolv and Tabu search
\end{itemize}
}

Correlation between Hardware and Software \redHL{H\"usrev}

Once the optimization problem is formulated in the software, the problem can be mapped to the hardware. Hardware mapping first needs to ensure that the formulation is in Ising form; if the problem is formulated in QUBO, then transformation identities can be used to translate the QUBO matrix to the Ising matrix and local field coefficients. Another requirement is to round the coefficients, as the digital RO implementation only supports integer coupling values in $[-14,+14]$ range. However, even if the original problem is formulated as integers, the desired coupling weights might not fit in the device range. Therefore the weight values are needed to be scaled. Performing scaling under the ceiling function ensures that even the smallest coupling values will be protected. The device accuracy is proportional to the coupling strength and therefore large scaling values are preferable. Whenever the magnitude of the coupling weights is stronger than the device's dynamic range, such weights have to be truncated to the minimum/maximum coupling value of the device. Whenever the truncation is performed excessively, the coupling matrix can be altered too much to represent the original problem energy space -- effectively introducing new global minima. Therefore scaling introduces trade-offs that must be carefully addressed, and dynamically reconfiguring the weight resolution of each spin allows better utilization of the hardware. 

A similar trade-off exists for the local field oscillators. The device allows an arbitrary number of spins to be configured as local field ROs (which implements $h$), while the remaining spins are configured to maintain pairwise coupling ($J$ values). Increasing the number of Local Field Ring Oscillators (LFROSCs) can increase the dynamic range of the $h$ coefficients in practice. If the effective dynamic range of the $h$ is more than $J$, then increasing LFROSC count can allow higher scaling values with less truncation, improving the accuracy dramatically. However, this also translates into less number of spin variables available for a subproblem mapping, causing an accuracy loss when a large problem is decomposed to problems that are small enough to fit in the device.
}
\section{Workflow of our Hybrid Approach}
\label{sec:Workflow}

\subsection{Overview of our Hybrid Solver Approach}
\label{ssec:hybrid_solver}

\noindent
Any hardware solver is limited in the number of spins that it can represent. Larger problems must be decomposed into smaller subproblems that can fit on the hardware and solved iteratively until the ground state is found.  The qbsolv~\cite{qbsolv} engine performs a similar decomposition, purely in software, optimizing a large QUBO problem by solving a series of sub-QUBO problems using local Tabu search. 

\begin{figure}[tb]
\vspace{0em}
\centering
\includegraphics[width=3.5in]{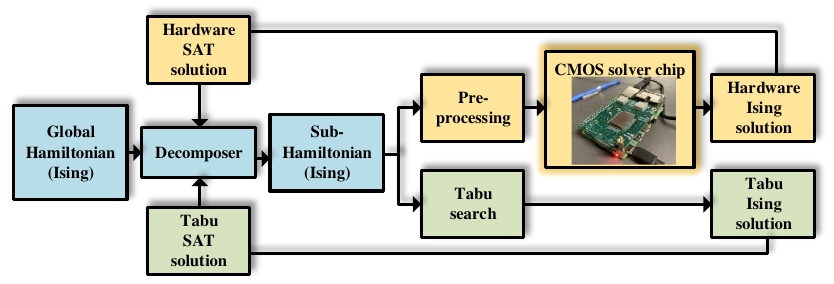}
\vspace{-6mm}
\caption{Workflow of our hybrid solver.}
\label{fig:workflow}
\vspace{-6mm}
\end{figure}

Our workflow for 3SAT solution and evaluation is illustrated in Fig.~\ref{fig:workflow}.  We show a hybrid hardware-based flow and a purely software-based flow. The software flow is based on qbsolv, but augmented with new methods that we propose for problem decomposition. 

The software approach effectively emulates the A2A hardware results, but is free of nonidealities and noise associated with a hardware implementation of an Ising solver, and should represent the best achievable results for the hardware-based flow.  Therefore, we use the software flow to evaluate the effectiveness of various Ising formulations (Section~\ref{sec:Formulations}) and decomposer schemes (Section~\ref{ssec:decomposition}). We evaluate this pruned list of candidates on the Ising hardware. The advantage of this two-step approach is that the hardware, which is an experimental university project, currently focuses on optimizing the core A2A engine but has slow I/O. This makes hardware evaluation of all algorithms and decomposers time-intensive.  This I/O bottleneck is a temporary artifact and is easily resolved: future versions of the chip, currently in the tapeout stage, are addressing this issue~\cite{Kim-private}.

In the figure, the blue boxes represent segments common to both solutions, and are performed on a classical computer. We first transform the 3SAT problem, consisting of variables and clauses, into a global Hamiltonian (Ising) formulation, as described in Section~\ref{sec:qubo_ising}. The size of the Hamiltonian depends on the 3SAT problem and the specific formulation used to map 3SAT to QUBO, as described in Section~\ref{sec:Formulations}. Typically, the Ising global Hamiltonian involves a large number of spins that cannot be directly mapped onto an on-chip RO array. To address this, we use a decomposer to generate an Ising sub-Hamiltonian that can fit the dimensions of the RO array: on the A2A hardware, this allows at most 49 spins, including the reference. 

For {\bf hardware-based} evaluation (yellow boxes in Fig.~\ref{fig:workflow}), this sub-Hamiltonian is taken through a preprocessing step (Section~\ref{sec:RO-preprocessing}), and the Ising weights are then programmed on to the RO-based Ising chip. The spin solution on the chip is sampled to determine the phase for each RO using majority voting. This represents the Ising solution for the sub-Hamiltonian, from which the SAT solution is inferred. 
To assess the accuracy of our RO solver, we also utilize a {\bf software-based} Tabu search~\cite{Glover1998}, as in qbsolv~\cite{qbsolv}, to solve the same sub-Hamiltonian as a control group (green boxes in Fig.~\ref{fig:workflow}) to find a Tabu spin solution from which we infer the SAT solution.  

For both hardware-based and software-based evaluation, after the solution of the current sub-Hamiltonian, the decomposer sends the next sub-Hamiltonian for processing.  This terminates after checking whether all clauses are satisfied, i.e., ``all-SAT'' is achieved, or if a predefined iteration limit is reached.  When a single SAT variable is determined by multiple spins (e.g., MIS$^{3m}$), it is crucial to identify and report contradictions when converting the Ising solution back to the SAT solution. Section~\ref{sec:Experiments} will delve deeper into these issues.

\subsection{Decomposers}
\label{ssec:decomposition}

\noindent
We study five decomposers: the last three are developed by us.

\noindent
{\bf Energy impact decomposer.} 
This decomposer, the qbsolv default, arranges all spins in ascending order based on their flip energy (i.e., the energy difference when spin $s_{i}$ is flipped to $-s_{i}$), and then selects spins with the highest flip energy to construct the sub-Hamiltonian.
However, such a greedy algorithm has a higher probability of being trapped in a local minimum where all spins have positive flip energy. 

\noindent
{\bf Random decomposer.} 
This decomposer randomly selects spins from the global Hamiltonian to form the sub-Hamiltonian. The randomness in the selection process helps the algorithm escape from local minima.

\noindent
{\bf Pseudorandom decomposer.}
Our heuristic scheme continually reshuffles and shifts the variable order, selecting the first $S$ variables at each iteration for the sub-Hamiltonian solver, where $S$ represents the number of spins supported by the hardware. The pseudorandom decomposer offers a lower probability of selecting the same spins in subsequent iterations than the Random decomposer, resulting in greater diversity among the generated sub-Hamiltonians.

\noindent
{\bf BFS decomposer.}  This scheme creates a cluster around a randomly-selected source vertex in the Ising graph. A breadth-first search (BFS) in the graph starts from the source, adding neighbors of vertices in randomized order until the sub-problem capacity is reached.

\noindent
{\bf SAT decomposer.}
This clustering-based decomposer randomly selects one clause at each step and adds all spins related to the selected clause and its involved variables to the sub-Hamiltonian. 

\ignore{
For example, assuming the clause decomposer randomly selects clause~1 as $(x_{1}\lor x'_{2}\lor x_{4})$, it adds $x_{1}$, $x'_{2}$ and $x_{4}$ to the cluster. 
In the MIS$^{3m}$ formulation, spins associated withe the variables $x_{1}$, $x'_{2}$ and $x_{4}$ in clause~1 are added; 
in the $\text{Chancellor}^{n{+}m}$ formulation, a clause-based spin for clause 1, and variable-based spins for $x_{1}$, $x_{2}$ (recall that $x'_{2}$ is represented in terms of spin $x_2$ in this formulation), and $x_{4}$ are added; and in the $\text{N\"u{\ss}lein}^{2n{+}m}$ formulation, a clause-based spin for clause~1, and two variable-based spins for each of $x_{1}$, $x'_{2}$, and $x_{4}$ are added.}


\subsection{Preprocessing for RO Array}
\label{sec:RO-preprocessing}

\noindent
The decomposed sub-Hamiltonians generated above must be further processed to work within the restrictions imposed by the Ising chip:
\begin{enumerate}
\item 
Coupling values must be integers in $[-14,+14]$; couplings for 3SAT formulations are integers, but may go out of this range.
\item 
Empirically, it is seen that the device accuracy is proportional to the coupling strength, and large coupling values are preferable. 
\item 
Empirically, device accuracy decreases for lower graph density. 
\end{enumerate}
Therefore, we preprocess the Ising Hamiltonian, translating the original sub-Hamiltonian from the decomposer to a hardware-compatible sub-Hamiltonian coupling matrix, using the following methods (these are applied to the hardware flow, not the software flow):

\begin{figure}[tb]
\vspace{0em}
\centering
\includegraphics[width=1.8in]{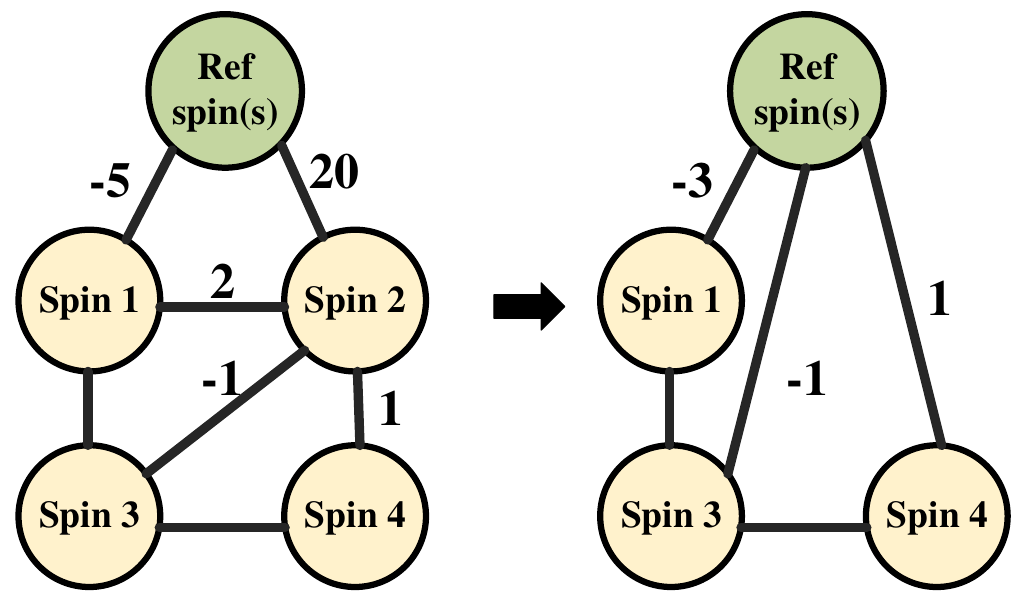}
\vspace{0em}
\caption{An example where spin~2 can be removed (set to the reference spin).}
\label{fig:remove}
\vspace{-6mm}
\end{figure}

\noindent
\textbf{Mapping}: As mentioned in Section~\ref{ssec:mapping_Ising}, increasing the number of LFROs and using spin merging to implement the local field can increase the dynamic range of the $h$ coefficients, to improve the solving accuracy. In Section~\ref{sec:Results}, we sweep the number of LFROs 
to locate the optimal number of LFROs to obtain an empirical value for the number of LFROs to be merged.

\noindent
\textbf{Removing spin variables from the sub-Hamiltonian}: When the local field on a spin variable is very large, it will force the variable to a fixed value. The contribution of a spin $s_i$ on the Ising Hamiltonian is $H(s_i) = h_i s_i + \textstyle \sum_{j=1, j \not = i}^n J_{ij} s_i s_j$. Then the flip energy\\
$H_{flip}(s_i) = H(s_i$$=$$+1) - H(s_i$$=$$-1) = 2 \left ( h_i + \textstyle \sum_{j=1, j \neq i}^n J_{ij} s_j \right )$\\
A lower bound smallest value of the second term is $\textstyle -\sum_{j=1, \\ j \neq i}^n |J_{ij}|$. Thus, if $h_i > 0$, then regardless of the choice of the other spins,
\begin{align}
H_{flip}(s_i) \geq 2 \left(h_i - \textstyle \sum_{j=1, j \neq i}^n |J_{ij}|\right)
\end{align}
If $h_i > \sum_{j=1, j \neq i}^n |J_{ij}|$, then $H(s_i = +1) > H(s_i = -1)$, i.e., a minimum Hamiltonian will force $s_i = -1$. Similarly, if $H_{flip}(s_i) \leq 2 (h_i + \sum_{j=1, j \not = i}^n |J_{ij}|) < 0$, i.e., $h_i < -\sum_{j=1, j \not = i}^n |J_{ij}|$, then it can be shown that $s_i = +1$ at a minimum Hamiltonian value. 

Together, these cases show that {\bf if $h_i$ is very negative [very positive], $s_i$ must be $+1$ [$-1$] at the minimum}, and the corresponding spin variable can be removed from the Hamiltonian.  Since the weights in the hardware are limited to $[-14,+14]$, this serendipitously allows us to remove large/small weights, at no loss of accuracy.

In practice, we use the criterion $|h_i|> N \times \max_j |J_{ij}|$, for a tuned value of $N$, and find it to be effective in identifying spins that could be removed, empirically without loss of optimality. In Fig.~\ref{fig:remove}, the coupling between spin~2 and reference spin~(Ref) $h_2=20$ is much larger than $\max_j J_{2j} = 2$. Therefore, spin~2 is removed from the Hamiltonian. All couplings, $J_{2j}$, are transferred into couplings with the reference spin~($h_j$), and spin~2 is set to $+1$.

\ignore{\redHL{Original text, to be removed}
In some scenarios, we can directly assign phase to a specific spin $S_i$ based on corresponding $h_i$ and $J_{ij}$, and then remove the spin from the implemented sub-Hamiltonian. After setting up a threshold value of $N$, if $|h_i|>N\times |J_{ij}|$ is true for a specific spin $S_i$, and any other spin $j$, there is a high probability that $S_i$ is dominated by $h_i$. Therefore, we can directly assign the phase for $S_i$ based on the sign of coupling $h_i$. For the example in Fig.~\ref{fig:remove}, the coupling $h_2$ is much higher than other $J_{2j}$, then spin~2 is removed from the Hamiltonian; all couplings with it~($J_{ij}$) are transferred into couplings with the reference spin $h_j$. Removing can reduce the number of ROs required on the hardware without obvious accuracy decreasing. 
}

\noindent
\textbf{Scaling}: Since large coupling values are preferable for device accuracy, and minimizing $F(\mathbf{s})$ in \eqref{eq:ising} is identical to minimizing $kF(\mathbf{s})$ for any scalar $k$, we can scale the $h_i$ and $J_{ij}$ values up as long as the maximum value lies in the range $[-14,+14]$. 
If a few spins go slightly above $\pm 14$, these may be truncated, as described next. 

\noindent
\textbf{Truncation}: Coupling values beyond $\pm 14$ can be truncated by clamping them to $=14$ or $-14$, whichever is closer. 


\ignore{
\subsection{Sampling}
\label{sec:sampling}

\noindent
The Ising chip generates 8 sampled bits for each ring oscillator (RO), indicating its phase. To determine the phase for each RO, we utilize a {\em majority voting} strategy. The majority voting process is based on the comparison of ``1'' and ``0'' occurrences in the 8-bit binary sample. We divide the samples into two groups: those with greater than or equal to half of the bits being ``1'' (4 or more out of 8), and those with less than half of the bits being ``1''. In the initial step, the sampler examines the sample of the reference spin(s) using this criterion, as all other spins will be compared to the reference spin. 
Then, all other spins that belong to the same side as the reference spin(s) are assumed to be in-phase, while the spins on the opposite side are considered out-of-phase. Based on their determined state (in-phase or out-of-phase), the sampler assigns a value of ``$+1$'' for the in-phase spins, and a value of ``$-1$''for those out-of-phase spins, which is the solution of sub-Hamiltonian.
}
\section{Experimental Setup and Metrics}
\label{sec:Experiments}

\noindent
We present three experiments in this section: (1)~a software simulation, focusing on Hamiltonian formulations and choices for the decomposer; (2)~a hardware sub-Hamiltonian test, focusing on preprocessing; (3)~using the optimal configuration from the first two experiments to solve 3SAT~benchmarks using our complete workflow. 

We use 10 benchmarks from the SATLIB uf20-91 suite~\cite{uf20-91}. All benchmarks are satisfiable and each includes 20 variables/91 clauses. With a clauses-to-variables ratio, $(m/n)$ of 4.55, these problems lie close to the \emph{phase transition region}~\cite{mitchellSAT1992}, where
about half of the SAT instances are satisfiable and the rest are unsatisfiable: the hardest SAT problems lie in this region.  We use the following metrics:

\noindent\textbf{Iterations}: In each {\em iteration} in the inner loop of our workflow, the decomposer generates a sub-Hamiltonian, sent to the Ising solver. The 3SAT problem is solved over multiple sub-Hamiltonian solutions.

\noindent\textbf{Repeats}: 3SAT/Hamiltonian solving is {\em repeated} many times in the outer loop of our workflow, thus reducing the impact of the random initial state 
of the chip, which may trap solutions in local minima.


\noindent\textbf{All-SAT rate}: This quality metric is the number of repeats that find an all-SAT solution, divided by the total number of repeats.

\noindent\textbf{Energy rate}: This is the ratio of the sub-Hamiltonian energy of the current solution and the ground state (from the software workflow), and indicates the accuracy of the sub-Hamiltonian solution.

\ignore{
Motivation of Experiments
\begin{itemize}
\item  \textbf{Decomposer: Our decomposer can achieve higher clauses-SAT rate over the general dwave QUBO/Ising decomposers on SAT problems in the formulation of MIS and new ILP}
\begin{itemize}
\item \red{Benchmarks: uf-20-91(01-100), (uf-50-218 and uf-75?)}
\item Testbenchs: QUBO(Ising) formulations of SAT in MIS, new ILP
\item Software test: Hamiltonians, solution, energy, (clause-SAT rate, all-SAT rate)
\begin{itemize}
\item \red{Decomposers: Energy impact, Random, pseudo-random, SAT-clauses, (modified)BFS}
\item \red{Size of subproblems: ratio of global/sub-problems}
\end{itemize}
\end{itemize}

\item  \textbf{Subproblems: Our Hardware can achieve higher speed, higher energy efficiency over software solver and good accuracy on SAT subproblems in the formulation of MIS and new ILP}
\begin{itemize}

\item \red{Benchmarks: uf-20-91(01-100), (uf-50-218 and uf-75)}
\item Testbenchs: Subproblems (Sub-Hamiltonians) of SAT in MIS, new ILP
\item Hardware and software metrics: 

\begin{itemize}
\item Energy overhead (hardware), 
\item \red{Energy consumption of chip}
\item \red{Energy consumption of Qbsolve (Tabu search)}
\item \red{Runtime of Qbsolve}, 
\end{itemize}

\item Hardware runtime
\begin{itemize}
\item Runtime (assume no interface barrier) 
\item Time for high-speed interface (estimation)
\item RO synchronization and sampling
\item Find minimal energy
\end{itemize}
\item Hardware test setup:
\begin{itemize}
\item MIS and new ILP
\item RO mapping to local field
\item Scaling
\item Remove (parameters)
\item Repeats
\item Energy and runtime for above Hamiltonian preprocessing
\end{itemize}
\item Software test setup
\begin{itemize}
\item num of repeat
\item to be added
\end{itemize}

\end{itemize}

\item  \textbf{3SAT solving: Our overall 3SAT solving platform can achieve high efficiency/ performance over the general Dwave QUBO/Ising solver.}
\begin{itemize}
\item \red{Benchmarks: uf-20-91(01-100), (uf-50-218 and uf-75?)}
\item Testbenchs: Subproblems (Sub-Hamiltonians) of SAT in MIS, new ILP or only new ILP
\item Ddudu-based SAT solving platform (best decomposer + best mapping)
\item Dwave QBsolve algorithm based on CPU
\item runtime estimation
\begin{itemize}
\item Typical runtime: 20s for 100 sample
\item one benchmark~(3SAT problem) 100 iterations: 2000s (33min)
\item one benchmark~(3SAT problem) 100 iterations * 100 repeats: (200,000s) = 55.6 hrs
\item 10 benchmarks~(3SAT problems): 55.6 * 10 = 556 hrs = 23-25 days (max time for data collection)

kim 31 41 122,
\end{itemize}
\end{itemize}
\end{itemize}

20-91
100*500 one problem 

50-218
100*1000
}

\label{sec:Results}

Next, we first sweep through different formulation and decomposer combinations using software-based solvers in Section~\ref{sec:experiments:decomposer_formulation}. We adapt the D-Wave Hybrid solver and replace its internal decomposition algorithm with the candidate decomposers from Section~\ref{ssec:decomposition}, and determine the best-performing formulation+decomposer in an ``ideal'' environment. In Section~\ref{ssec:experiments:hardware_optimization}, we perform Ising accelerator hardware parameter sweeps to maximize the solution accuracy for sub-Hamiltonian Ising problem instances. Finally, we combine software and hardware optimizations in Section~\ref{ssec:experiments:3sat_solution} to solve 3SAT instances.

\subsection{Decomposer and Formulation Results} \label{sec:experiments:decomposer_formulation}

\begin{figure}[tb]
\vspace{0em}
\centering
\includegraphics[width=0.8\linewidth]{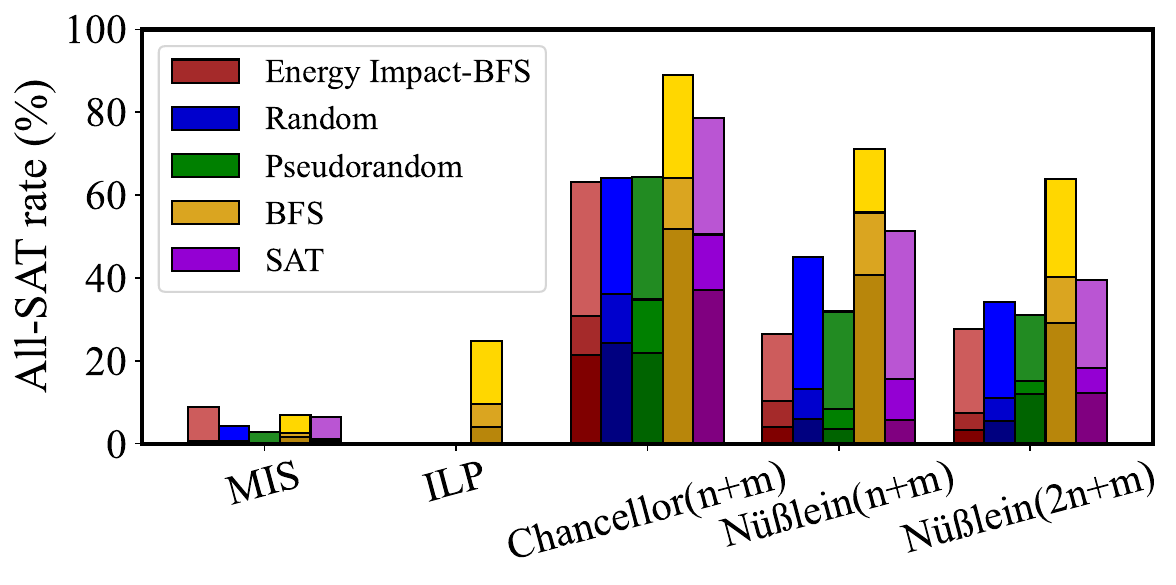}
\vspace{-4mm}
\caption{Evaluation of various formulations and decomposers (the pseudorandom, BFS, and SAT are developed in this work).}
\label{fig:formulations}
\vspace{-3mm}
\end{figure}

\noindent
In Fig.~\ref{fig:formulations}, we explore the different Ising formulations of the 3SAT problem alongside different decomposition strategies using software-only subsolvers (i.e., Tabu search). 
The figure shows the average All-SAT rate out of 100 repeats for the first 10 instances in the uf20-91 benchmarks, using multiple formulations and decomposers, over 500 iterations. The results show the All-SAT rate at the end of 50, 100, and 500 iterations, as denoted with darker-to-lighter tones from bottom to top in each bar.
\emph{We conclude that the best performance comes from the Chancellor\textsuperscript{$n{+}m$} formulation and the BFS decomposer.}

\ignore{
\begin{figure}[tb]
\definecolor{darkyellow}{HTML}{D0B22B}
    \centering
    \resizebox{0.7\linewidth}{!}{
    \begin{tikzpicture}
    \begin{axis}[
        cycle list={{blue},{red},{green!60!black},{black},{darkyellow}},
        axis line style={draw=none},
        tick style={draw=none},
        axis background/.style={fill=gray!5},
        width=.9\columnwidth,
        height=.65\columnwidth,
        grid=both, 
        grid style={white},
        legend cell align={left},
        legend entries={BFS, EnergyImpact-BFS, Pseudorandom, Random, SATdecomposer},
        legend style={fill==gray!10, fill opacity=0.4, draw opacity=1, text opacity=1, draw=gray!20},
        legend pos=south east,
        xlabel={Iteration number},
        ylabel={Average All-SAT rate (100 reps)},
        enlargelimits=false
    ]
        \addplot +[mark=none, thick] table {data/decomposer_subtabu3ms/comparison_uf20-91_BFS.txt}; 
        \addplot +[mark=none, thick] table {data/decomposer_subtabu3ms/comparison_uf20-91_EnergyImpact-BFS.txt};
        \addplot +[mark=none, thick] table {data/decomposer_subtabu3ms/comparison_uf20-91_Pseudorandom.txt};
        \addplot +[mark=none, thick] table {data/decomposer_subtabu3ms/comparison_uf20-91_Random.txt};
        \addplot +[mark=none, thick] table {data/decomposer_subtabu3ms/comparison_uf20-91_SATdecomposer.txt};
    \end{axis}
    \end{tikzpicture}
    }
    \vspace{-2mm}
    \caption{Average All-SAT rate up to 500 iterations, with 100 repeats, for the first 10 uf20-91 benchmarks using $\text{Chancellor}^{n{+}m}$ formulation.}
    \label{fig:decomposer_subtabu3ms}
    \vspace{-6mm}
\end{figure}

Under the same experimental settings, Fig.~\ref{fig:decomposer_subtabu3ms} shows the evolution of the All-SAT rate for the Chancellor$^{n+m}$ formulation for various decomposers.
Empirically, we see that BFS provides the most All-SAT solutions for the 20-variable benchmarks, followed by SATdecomposer, Pseudorandom decomposer, Random decomposer, and EnergyImpact decomposer with BFS traversal. The reason why the Pseudorandom decomposer outperforms the Random decomposer can be explained by the fact that it avoids the repetitive accumulation of the same variable groups in the same subproblem over time.
}


\subsection{Hardware Optimization Results} \label{ssec:experiments:hardware_optimization}

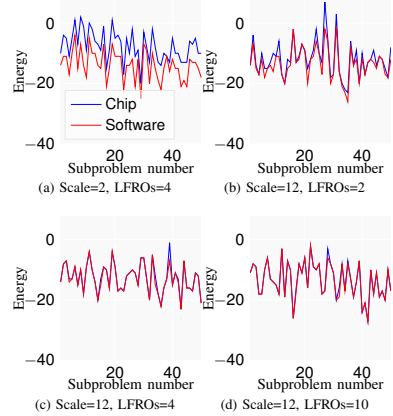
\begin{figure}[tb]
    \centering
    \resizebox{0.6\linewidth}{!}{
    \subfloat[Scale=2, LFROs=4]{
        \begin{tikzpicture}
        \begin{axis}[
            axis line style={draw=none},
            tick style={draw=none},
            axis background/.style={fill=gray!5},
            width=.49\columnwidth,
            height=.5\columnwidth,
            grid=both, 
            grid style={white},
            legend cell align={left},
            legend entries={\small{Chip},\small{Software}},
            legend style={fill==gray!10, fill opacity=0.6, draw opacity=1, text opacity=1, draw=gray!20},
            legend pos=south west,
            xlabel={\small{Subproblem number}\vspace{-0.3cm}},
            ylabel={\small{Energy}},
            ylabel style={yshift=-.28cm},
            xlabel style={yshift=.2cm},
            enlargelimits=false,
            ymax=8, ymin=-40
        ]
            \addplot +[mark=none] table {data/subproblem_accuracy_ground/chip_uf20-01-scale2-size45lfsize4.dat}; 
            \addplot +[mark=none] table {data/subproblem_accuracy_ground/qbsolv_uf20-01-scale2-size45lfsize4.dat}; 
        \end{axis}
        \end{tikzpicture}
    } \hspace{-.4cm}  
    \subfloat[Scale=12, LFROs=2]{
        \begin{tikzpicture}
        \begin{axis}[
            axis line style={draw=none},
            tick style={draw=none},
            axis background/.style={fill=gray!5},
            width=.49\columnwidth,
            height=.5\columnwidth,
            grid=both, 
            grid style={white},
            legend cell align={left},
            legend style={fill==gray!10, fill opacity=0.6, draw opacity=1, text opacity=1, draw=gray!20},
            legend pos=south west,
            xlabel={\small{Subproblem number}},
            ylabel={\small{Energy}},
            ylabel style={yshift=-.28cm},
            xlabel style={yshift=.2cm},
            enlargelimits=false,
            ymax=8, ymin=-40
        ]
            \addplot +[mark=none] table {data/subproblem_accuracy_ground/chip_uf20-01-scale12-size47lfsize2.dat}; 
            \addplot +[mark=none] table {data/subproblem_accuracy_ground/qbsolv_uf20-01-scale12-size47lfsize2.dat}; 
        \end{axis}
        \end{tikzpicture}
    }}\\
    \resizebox{0.6\linewidth}{!}{
    \vspace{-.35cm}
    \subfloat[Scale=12, LFROs=4]{
        \begin{tikzpicture}
        \begin{axis}[
            axis line style={draw=none},
            tick style={draw=none},
            axis background/.style={fill=gray!5},
            width=.49\columnwidth,
            height=.5\columnwidth,
            grid=both, 
            grid style={white},
            legend cell align={left},
            legend style={fill==gray!10, fill opacity=0.6, draw opacity=1, text opacity=1, draw=gray!20},
            legend pos=south west,
            xlabel={\small{Subproblem number}},
            ylabel={\small{Energy}},
            ylabel style={yshift=-.28cm},
            xlabel style={yshift=.2cm},
            enlargelimits=false,
            ymax=8, ymin=-40
        ]
            \addplot +[mark=none] table {data/subproblem_accuracy_ground/chip_uf20-01-scale12-size45lfsize4.dat}; 
            \addplot +[mark=none] table {data/subproblem_accuracy_ground/qbsolv_uf20-01-scale12-size45lfsize4.dat}; 
        \end{axis}
        \end{tikzpicture}
    }\hspace{-.4cm}
    \subfloat[Scale=12, LFROs=10]{
        \begin{tikzpicture}
        \begin{axis}[
            axis line style={draw=none},
            tick style={draw=none},
            axis background/.style={fill=gray!5},
            width=.49\columnwidth,
            height=.5\columnwidth,
            grid=both, 
            grid style={white},
            legend cell align={left},
            legend style={fill==gray!10, fill opacity=0.6, draw opacity=1, text opacity=1, draw=gray!20},
            legend pos=south west,
            xlabel={\small{Subproblem number}},
            ylabel={\small{Energy}},
            ylabel style={yshift=-.28cm},
            xlabel style={yshift=.2cm},
            enlargelimits=false,
            ymax=8, ymin=-40
        ]
            \addplot +[mark=none] table {data/subproblem_accuracy_ground/chip_uf20-01-scale12-size39lfsize10.dat}; 
            \addplot +[mark=none] table {data/subproblem_accuracy_ground/qbsolv_uf20-01-scale12-size39lfsize10.dat}; 
        \end{axis}
        \end{tikzpicture}
    }
    }
    \caption{Software and chip Hamiltonian energy for uf20-91/01 instance, (a)~for Scale~2 with 4~LFROs; for Scale~12 with (b)~2, (c)~4, and (d)~10~LFROs. 
    }
    \label{fig:subproblem_ground_scale_lfrosc}
    \vspace{-4mm}
\end{figure}

\noindent
In order to optimize the parameters that improve the subsolution accuracy on the hardware, we show the effects of scaling and mapping (i.e., number of LFROs) in Fig.~\ref{fig:subproblem_ground_scale_lfrosc}. Individual software and chip energies for the subproblems that are constructed through decomposition for uf20-91/01 benchmark are provided in Fig.~\ref{fig:subproblem_ground_scale_lfrosc}(a) for a scaling factor of 2 (i.e., all couplings are multiplied by 2) and in Fig.~\ref{fig:subproblem_ground_scale_lfrosc}(b)--(d) for scaling by 12 with 2, 4, and 10 LFROs, respectively.   Scaling is used together with truncation, clamping values beyond $\pm 14$. Note that Fig.~\ref{fig:subproblem_ground_scale_lfrosc}(a) has a significant discrepancy between the ideal and hardware Hamiltonian energies, while in Fig.~\ref{fig:subproblem_ground_scale_lfrosc}(c) with higher scaling and the same number of LFROs, the hardware follows the software much more closely due to the stronger coupling of spins, thus reinforcing the empirical observation that stronger coupling improves chip accuracy. At the higher scaling factor of 12, it is seen that 2~LFROs in Fig.~\ref{fig:subproblem_ground_scale_lfrosc}(b) provide better accuracy than the Scale~2 case, but that 4 and 10~LFROs in Fig.~\ref{fig:subproblem_ground_scale_lfrosc}(c) and (d), respectively, provide progressive improvements. This is because the larger number of LFROs have sufficient dynamic range to accommodate $h$ coefficients with fewer truncations. \emph{Therefore, appropriate selection of scaling factors and the LFROs can make hardware subproblem solutions more closely follow the software-based ground state solutions.}

\ignore{
Similarly, the subproblem energies are provided in Fig.~\ref{fig:subproblem_ground_scale_lfrosc}(c) for 2~LFROs and in Fig.~\ref{fig:subproblem_ground_scale_lfrosc}(d) for 10~LFROs
when the scaling factor is 12. As before, the smaller scaling factor has a slight discrepancy between the ideal and hardware Hamiltonian energies, while in Fig.~\ref{fig:subproblem_ground_scale_lfrosc}(d) the hardware follows the software much more closely due to the larger number of LFROs having a sufficient dynamic range to accommodate $h$ coefficients with less number of truncations.
\emph{Therefore, appropriate selection of scaling factors as well as the number of LFROs can indeed make hardware subproblem solutions more closely follow the software-based ground state solutions.}
}

We sweep the scaling factor and the number of LFROs, and based on the statistics of subproblem accuracy, we choose an optimal point of peak accuracy at a scaling factor of 12, and a choice of 4 LFROs.

\ignore{
\begin{figure}[tb]
    \centering
    \resizebox{0.85\linewidth}{!}{
    \subfloat[LFROs=4]{
        \begin{tikzpicture}
        \begin{axis}[
            axis line style={draw=none},
            tick style={draw=none},
            axis background/.style={fill=gray!5},
            width=.54\columnwidth,
            height=.5\columnwidth,
            grid=both, 
            grid style={white},
            xlabel={Scaling Factor},
            ylabel={\%Energy rate},
            enlargelimits=false,
            ymax=102,
            ymin=0,
            xmin=2,
            xmax=12,
            ylabel style={yshift=-.3cm},
            xlabel style={yshift=.16cm},
        ]
\addplot + table [x=scale,y=accuracy]{data/subproblem_accuracy/scale.dat}; 
\addplot [name path=upper,draw=none] table[x=scale,y expr=\thisrow{accuracy}+\thisrow{err}] {data/subproblem_accuracy/scale.dat};
\addplot [name path=lower,draw=none] table[x=scale,y expr=\thisrow{accuracy}-\thisrow{err}] {data/subproblem_accuracy/scale.dat};
\addplot [fill=blue!10] fill between[of=upper and lower];
\node[anchor=north west] at (axis cs:2,50   ) {\scriptsize{0\%}};
\node[anchor=north     ] at (axis cs:4.5,90 ) {\scriptsize{0\%}};
\node[anchor=north     ] at (axis cs:6,92   ) {\scriptsize{9\%}};
\node[anchor=north     ] at (axis cs:7.8,98)  {\scriptsize{51\%}};
\node[anchor=north     ] at (axis cs:9.6,99)  {\scriptsize{78\%}};
\node[anchor=north east] at (axis cs:12.4,98){\scriptsize{96\%}};

        \end{axis}
        \end{tikzpicture}
    }
    \hspace{-.7cm}
    \subfloat[Scale=12]{
        \begin{tikzpicture}
        \begin{axis}[
            axis line style={draw=none},
            tick style={draw=none},
            axis background/.style={fill=gray!5},
            width=.54\columnwidth,
            height=.5\columnwidth,
            grid=both, 
            grid style={white},
            xlabel={\#LFROs},
            ylabel={\%Energy rate},
            enlargelimits=false,
            ymax=102,
            ymin=0,
            xmin=2,
            xmax=12,
            ylabel style={yshift=-.3cm},
            xlabel style={yshift=.16cm},
        ]
\addplot + table [x=lfrosc,y=accuracy]{data/subproblem_accuracy/lfrosc.dat}; 
\addplot [name path=upper,draw=none] table[x=lfrosc,y expr=\thisrow{accuracy}+\thisrow{err}] {data/subproblem_accuracy/lfrosc.dat};
\addplot [name path=lower,draw=none] table[x=lfrosc,y expr=\thisrow{accuracy}-\thisrow{err}] {data/subproblem_accuracy/lfrosc.dat};
\addplot [fill=blue!10] fill between[of=upper and lower];
\node[anchor=north west] at (axis cs:1.6,88)  { \scriptsize{100\%}};
\node[anchor=north     ] at (axis cs:4,98)    { \scriptsize{96\%}};
\node[anchor=north     ] at (axis cs:5.9,100) { \scriptsize{81\%}};
\node[anchor=north     ] at (axis cs:7.8,98)  { \scriptsize{66\%}};
\node[anchor=north     ] at (axis cs:9.6,99) { \scriptsize{59\%}};
\node[anchor=north east] at (axis cs:12.4,98) { \scriptsize{54\%}};
        \end{axis}
        \end{tikzpicture}
    }
    }
    \caption{Average subproblem accuracy for varying (a) scaling (b) LFROs for LFROs=4 and scale=12, respectively, when solving the first 10 problems in uf20-91. The average ratio of truncated coefficients is annotated per data point.}
    \label{fig:subproblem_scale_lfrosc}
    \vspace{-6mm}
\end{figure}

Next, we examine the statistics of subproblem accuracy.
We vary the scaling factor in Fig.~\ref{fig:subproblem_scale_lfrosc}(a) fixing the number LFROs to 4, and in Fig.~\ref{fig:subproblem_scale_lfrosc}(b), we vary the number of LFROs, fixing the scaling factor to 12. We use truncation with scaling, and annotate each data point with the number of truncated coefficients.  The results are derived from the first 10 problems of uf20-91 SAT benchmarks for 50 iterations. In each figure, the solid line represents the mean of the Hamiltonian energy, and the shaded region marks one standard deviation from the mean, over all repeats.  In Fig.~\ref{fig:subproblem_scale_lfrosc}(a), until truncation alters the problem, increasing the coupling strength improves accuracy. In Fig.~\ref{fig:subproblem_scale_lfrosc}(b), more LFROs provide a sufficient dynamic range to accommodate local field coefficients with higher magnitude, until truncation degrades the solution, although the accuracy impact is not as dramatic as sweeping through the coupling strength in Fig.~\ref{fig:subproblem_scale_lfrosc}(a). 

We also see that both scaling factor and 
LFRO
optimizations affect the number of coupling weights that are truncated due to the insufficient dynamic range. A scaling factor of 2 provides the fewest truncated variables, but subproblem accuracy is low due to insufficient coupling strength. As the scaling factor increases, the subproblem accuracy improves despite more and more variables being truncated, showing the tradeoff between the dynamic range and the coupling strength. \emph{Thus, the energy landscape transformation due to insufficient dynamic range affects the subproblem solution accuracy in such a way that number of truncated coefficients is correlated with the accuracy loss.
Overall a scaling factor of 8 alongside 4 LFROs provides the highest subproblem accuracy in terms of \%Energy rate.}
}

\subsection{3SAT~Results with Ising Accelerator} \label{ssec:experiments:3sat_solution}




\begin{figure}[t]
\vspace{-6mm}
\centering
\resizebox{1.05\linewidth}{!}{\hspace{-7mm}
\subfloat[]{
\includegraphics[width=0.54\linewidth]{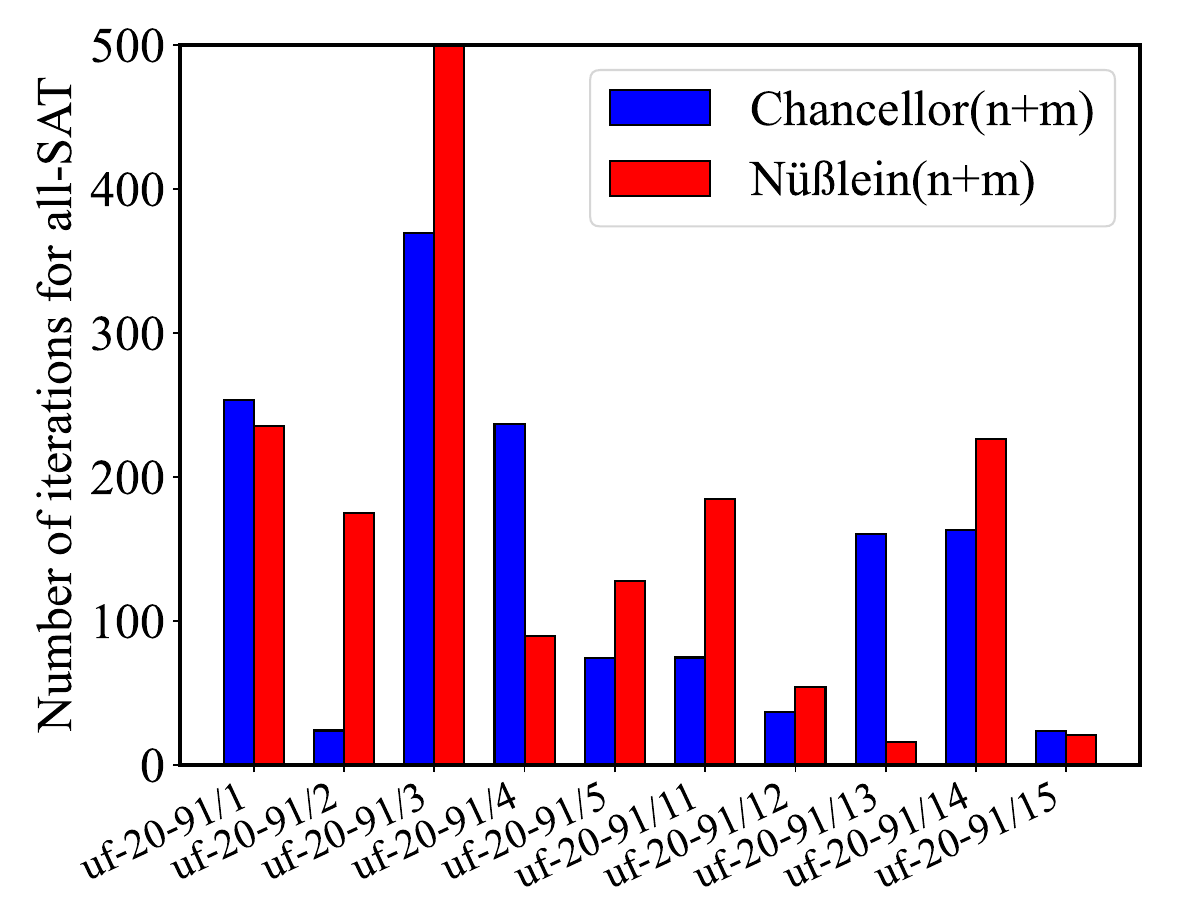}
}\hspace{-5mm}
\subfloat[]{
\includegraphics[width=0.54\linewidth]{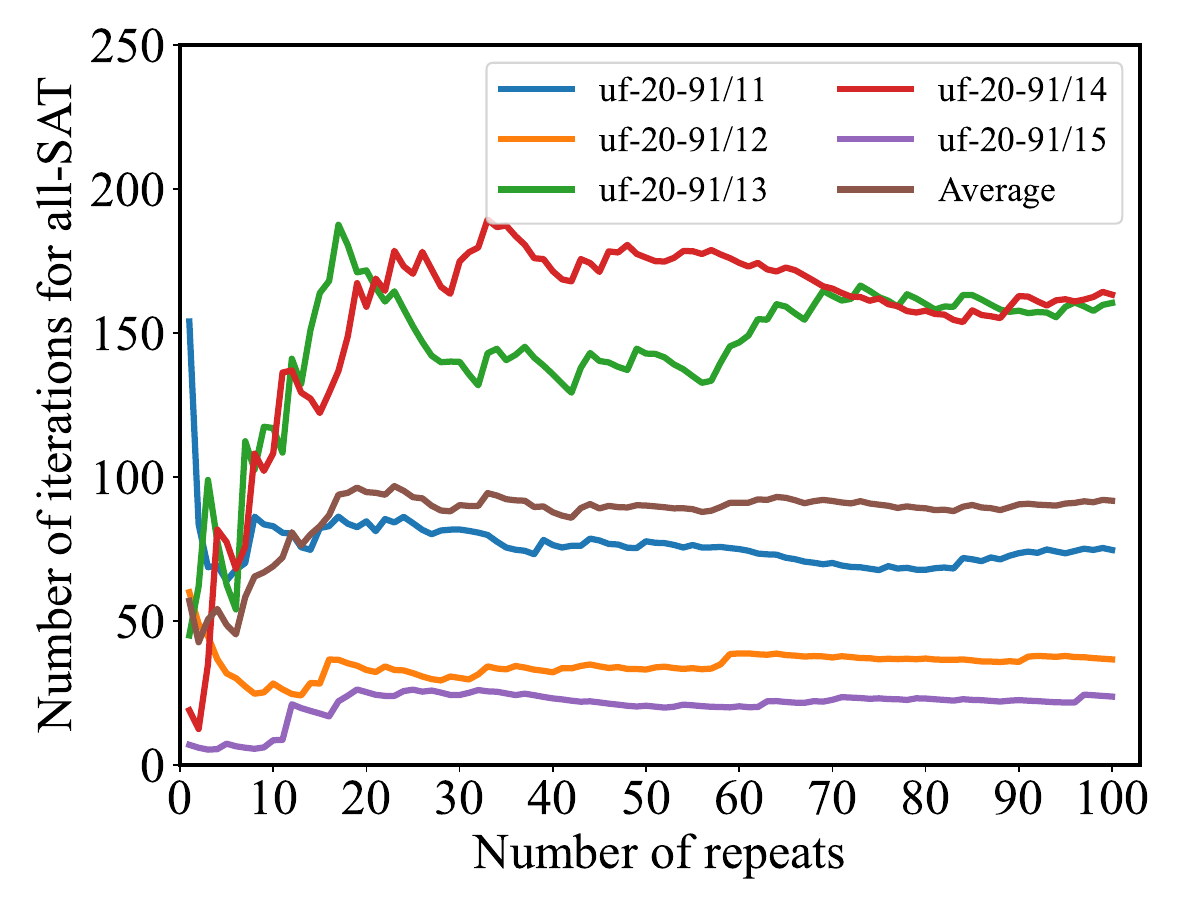}
}
}
\vspace{-2mm}
\caption{Number of iterations to find All-SAT for hardware test on the Ising chip for (a) Average number of iterations for uf20-91/(01-05, 11-15) for Chancellor\textsuperscript{$n{+}m$} and N\"u{\ss}lein\textsuperscript{$n{+}m$} (b) Average number of iterations with repeats for uf20-91/(11--15) in the formulation of Chancellor\textsuperscript{$n{+}m$}.}
\label{fig:hardware_results}
\vspace{-3mm}
\end{figure}

\noindent
Based on Fig.~\ref{fig:formulations}, we select the best two formulations (Chancellor\textsuperscript{$n{+}m$} and N\"u{\ss}lein\textsuperscript{$n{+}m$}) and use their best decomposer~(BFS) to setup our Hybrid 3SAT solver. We use instances 01--05 of uf20-91, which were used in our previous optimization, and an orthogonal set, instances 11--15 of uf20-91, which were never used in any previous experiments to optimize our hybrid solver. Over 100 repeats for each benchmark, with a time-out at 500 iterations, we plot the average number of iterations to reach all-SAT in Fig.~\ref{fig:hardware_results}(a) for the Chancellor\textsuperscript{${n{+}m}$} and N\"u{\ss}lein\textsuperscript{$n{+}m$} formulations, over all 10 instances.  In general, the Chancellor formulation is better as it achieves all-SAT in fewer iterations than N\"u{\ss}lein, with a few exceptions, e.g., uf20-91/13.  However, in one case (uf20-91/03), the N\"u{\ss}lein formulation is unable to achieve all-SAT over the 100 repeats, as shown by the average number of iterations being at the time-out limit of 500.
Fig.~\ref{fig:hardware_results}(b) shows the evolution of the average number of iterations over 100 repeats for instances 11--15. After some repeats, the average settles to the steady-state value in Fig.~\ref{fig:hardware_results}(a).

\begin{table}[t]
\caption{Runtime estimation for our Hybrid solver.}
\label{tab:runtime}
\vspace{-2mm}
\centering
\resizebox{\linewidth}{!}{
\begin{tabular}{|l|c|c|c|c|}
\hline
                       & Runtime per                     & Multiplier for  & Number of & Runtime       \\
                       & unit operation                  & previous column & repeats   & (ms)          \\ \hline
Input                  & $1.25\times 10^{-3}\mu$s/bit    & 9604 bits       & 1         & 12.0$\mu$s    \\ \hline
RO relaxation          & (1/26MHz) s/cycle               & 40 cycles       & 100       & 153.8$\mu$s   \\ \hline
Output                 & $1.25\times 10^{-3}\mu$s/bit    & 49 bits         & 100       & 6.1$\mu$s     \\ \hline \hline
\multicolumn{4}{|l|}{\bf Overall runtime}                                              & \textbf{171.9$\mu$s}  \\ \hline
\end{tabular}
}
\vspace{-5mm}
\end{table}

\noindent
{\bf Runtime analysis.}
A runtime analysis of our Hybrid solver is illustrated in Table~\ref{tab:runtime}, assuming 100 repeats and 100 iterations per repeat and finds the one with minimal Hamiltonian, which is consistent with Fig.~\ref{fig:formulations}.   
As an academic demonstrator, with a focus on optimizing the all-to-all array, the Ising chip has several limitations:\\
(1) It has low IO bandwidth -- but this is not a fundamental bottleneck. The authors of~\cite{Lo2023} indicate that they can achieve 800Mbps in a future version of the chip~\cite{Kim-private} with 8-bit parallel IO at 100Mbps each.\\
(2) Majority voting (Section~\ref{sec:cmos_hardware}) is performed externally on a CPU, but can easily be incorporated in a small on-chip circuit. With this on-chip voting engine, the output data can be reduced from 8 bits/spin to 1 bit/spin, so that 49 spins only generate 49 bits of output.\\
(3) The Hamiltonian computation to check for convergence is currently performed off-chip, using the Ising energy computation function in the dimod~\cite{dimod} package on an Intel Xeon 4114 CPU.  This currently takes 3.6ms, and in a currently taped-out advanced version of the Ising chip, this computation is performed in hardware~\cite{Kim-private}. The on-chip Hamiltonian computation is overlapped with the next Ising solution, and therefore incurs no additional latency.\\
All of these are relatively simple extensions, and the only reason that they are not on the chip already is because it is an academic project, limited by the number of students who can work on tape-outs; nevertheless, these enhancements are already in the pipeline in chips that are taped out or will be taped out shortly~\cite{Kim-private}.  To project the true power of this hardware computational model, we use the settling time of the current version of the Ising chip, and project the total runtime numbers under the assumption that the above three improvements are made.
Under these assumptions, the overall runtime is 171.9$\mu$s. Given that the software Tabu search usually takes 10-100ms, and the default timeout value is 100ms in D-Wave Hybrid~\cite{dwave-hybrid}, our hybrid solver can solve 3SAT problems orders of magnitude faster than a software-based solver. A preliminary experiment on a small 3SAT instance, which can fit on the Ising chip without decomposition, shows a 10$\times$ runtime improvement over a classical SAT solver, MiniSAT~\cite{minisat}.

\ignore{
\item Fig:5 Subproblems
\begin{itemize}
\item metrics: Minimal energy
\item use the best configuration above, compare the best of 100 samples with QBsolve
\item benchmarks: 10 benchmarks * 100 subproblems
\item time = 20*10*100 = 20000s = 5.6h
\end{itemize}

\item Fig:5 3SAT accuracy
\begin{itemize}
\item metrics: All-SAT rate, (clause SAT rate)
\item Qbsolve vs Ddudu
\item benchmarks: 10 benchmarks * 100 repeats * 100 iterations
\item time: 100,000 *20s = 556h = 23-24 days = 12 days with 2 boards
\end{itemize} 

\item Fig:6 3SAT runtime and energy vs software
\begin{itemize}
\item metrics: All-SAT rate, (clause SAT rate)
\item Qbsolve vs Ddudu
\item runtime: RO synchronization time vs tabu search time
\item energy: estimate the CPU energy consumption vs RO array energy
\item time: NA
\end{itemize}
}

\section{Conclusion}

\noindent
This work solves 3SAT on a CMOS-based Ising chip, addressing degrees of freedom in problem formulation and problem decomposition, as well as hardware mapping strategies for the Ising problem that extract the best performance from the chip. To the best of our knowledge, this is the first comprehensive exploration of these issues,
paving
the way towards bringing Ising computation to the mainstream through algorithm mapping on a mass-manufacturable CMOS chip.  

\bibliographystyle{misc/IEEEtran}
\bibliography{bib/main}
\end{document}